\documentclass{ita}
\usepackage[latin1]{inputenc}
\usepackage{amssymb}

\newtheorem{The}{Theorem}[section]
\newtheorem{Pro}[The]{Proposition}
\newtheorem{Deff}[The]{Definition}
\newtheorem{Lem}[The]{Lemma}
\newtheorem{Rem}[The]{Remark}

\newtheorem{Cor}[The]{Corollary}
\newtheorem{Not}[The]{Notation}

\newcommand{\fa}{\forall}
\newcommand{\Ga}{\Gamma}

\newcommand{\Gao}{\Gamma^\omega}

\newcommand{\Si}{\Sigma}
\newcommand{\Sis}{\Sigma^\star}
\newcommand{\Sio}{\Sigma^\omega}
\newcommand{\ra}{\rightarrow}
\newcommand{\hs}{\hspace{12mm}

\noi}

\newcommand{\ite}{\item}

\newcommand{\ol}{ $\omega$-language}

\newcommand{\om}{\omega}
\newcommand{\nl}{\newline}
\newcommand{\noi}{\noindent}

\newcommand{\proo}{\noi {\bf Proof.} }
\newcommand {\ep}{\hfill $\square$}

\begin{document}

\title{{\bf   Highly Undecidable Problems  \\ For  Infinite Computations}}

\runningtitle{Highly Undecidable Problems  for  Infinite Computations}
\author{Olivier Finkel}
\address{Equipe de Logique Mathématique
 \\  CNRS et Université Paris 7, France. \\ 
\email{finkel@logique.jussieu.fr }}

\date{}



\begin{abstract}
\noi We show that many   classical decision  problems 
 about   $1$-counter  $\om$-languages, context free  $\om$-languages, or infinitary rational relations, are 
 $\Pi_2^1$-complete, hence located  at the second level of the analytical hierarchy, and ``highly undecidable". 
In particular,  the  universality 
problem, the inclusion problem, the equivalence problem, the determinizability problem, the complementability problem, and the 
unambiguity problem  are all $\Pi_2^1$-complete for context-free $\om$-languages or for infinitary rational relations. Topological and arithmetical properties of 
$1$-counter  $\om$-languages, context free  $\om$-languages, or infinitary rational relations, are also highly undecidable. 
These very surprising results  provide the first examples of highly undecidable problems about the behaviour of very 
simple finite machines like $1$-counter automata or $2$-tape 
automata. 
\end{abstract}

\maketitle

\section{Introduction}

Many classical decision problems arise  naturally in the fields of  Formal Language Theory and of Automata Theory. 
When languages of finite words are considered it is well known that most problems about regular languages accepted by finite automata are decidable. 
On the other hand, at the second level of the Chomsky Hierarchy, most problems about context-free languages accepted by pushdown automata or generated by 
context-free grammars are undecidable. For instance it follows from the undecidability of the Post Correspondence Problem 
that  the universality problem, the inclusion and the equivalence problems for context-free languages are also undecidable. 
Notice that some few problems about context-free languages remain decidable like the following ones: ``Is a given  context-free language $L$  empty ? " 
``Is a given  context-free language $L$ infinite ? " ``Does a given word $x$ belong to a given  context-free language $L$  ? " 
Sénizergues proved  in \cite{Senizergues01}  that  the difficult problem of the equivalence of two deterministic pushdown automata is 
decidable. Another  problem about finite simple machines 
is the equivalence problem for deterministic multitape automata. It has  been proved to be decidable by Harju and Karhum\"aki in \cite{HarjuKarhumaki91}. 
But all known problems about acceptance by Turing machines  are undecidable, \cite{HopcroftMotwaniUllman2001}. 
\nl  Languages of infinite words accepted by finite automata were first studied by B\"uchi 
to prove the decidability of the monadic second order theory of one successor
over the integers.  Since then regular $\om$-languages have been much studied and  many applications have been  found  for specification and verification 
of non-terminating systems, 
see \cite{Thomas90,Staiger97,PerrinPin} for many results and references. 
More powerful  machines, like 
pushdown automata, Turing machines, 
 have also been considered  for the reading of infinite words, 
see Staiger's survey \cite{Staiger97}  and the fundamental study \cite{eh}
of Engelfriet and  Hoogeboom on {\bf X}-automata, i.e. finite automata equipped with 
a storage type {\bf X}. As in the case of finite words, most problems about regular  $\om$-languages have been shown to be decidable. On the other hand 
most problems about context-free $\om$-languages are known to be undecidable, \cite{CG}. 
Notice that almost all undecidability proofs rely on the undecidability of the Post Correspondence Problem which is complete for the class of recursively 
enumerable problems, i.e. complete at the first level of the arithmetical hierarchy. Thus undecidability results about context-free $\om$-languages provided 
only hardness results for the first level of the arithmetical hierarchy.
\nl Castro and Cucker studied decision problems for $\om$-languages of Turing machines in \cite{cc}. They studied the degrees
 of many classical decision problems like : ``Is the $\om$-language recognized by a given machine non empty ?", ``Is it finite ?" ``Do two given machines 
recognize the same $\om$-language ?"
\nl Their motivation was on one side to classify the problems about Turing machines and on the other side to ``give natural complete problems for the 
lowest levels of the analytical hierarchy which constitute an analog of the classical complete problems given in recursion theory for the arithmetical hierarchy". 
\nl On the other hand we showed in \cite{Fin-mscs06} that 
context free $\om$-languages, or even  $\om$-languages accepted  by B\"uchi $1$-counter automata, have the same topological complexity as  
  $\om$-languages accepted by Turing machines with a B\"uchi  acceptance condition. We use in this paper several constructions of  \cite{Fin-mscs06}
to infer some undecidability results from those of  \cite{cc}. Notice that one cannot infer directly from  topological results of  \cite{Fin-mscs06} that 
the degrees of decision problems for $\om$-languages of B\"uchi $1$-counter automata are the same as the degrees of the corresponding decision problems 
about Turing machines. For instance the non-emptiness problem and the infiniteness problem are decidable for 
$\om$-languages accepted by B\"uchi $1$-counter automata or even by B\"uchi pushdown automata but the non-emptiness problem and  
 the infiniteness problem for $\om$-languages of Turing machines are both $\Si_1^1$-complete, hence highly undecidable, \cite{cc}. 
However we can show that many other   classical decision  problems 
 about   $1$-counter  $\om$-languages or context free  $\om$-languages,  are 
 $\Pi_2^1$-complete, hence located  at the second level of the analytical hierarchy, and ``highly undecidable". 
In particular,  the  universality 
problem, the inclusion problem, the equivalence problem, the determinizability problem, the complementability problem, and the 
unambiguity problem  are all $\Pi_2^1$-complete for $\om$-languages of  B\"uchi $1$-counter automata. Topological and arithmetical properties of 
$1$-counter  $\om$-languages and of  context free  $\om$-languages are also highly undecidable. 
\nl  In another paper we had also shown that infinitary rational relations accepted by $2$-tape B\"uchi 
automata have the same topological complexity as  
  $\om$-languages accepted by B\"uchi $1$-counter automata or by B\"uchi Turing machines. 
This very surprising result was obtained by using a simulation of the behaviour of real time 
$1$-counter automata by $2$-tape B\"uchi automata,  \cite{Fin06b}. 
Using some  constructions of  \cite{Fin06b} we infer from results about degrees of decision problems for B\"uchi $1$-counter automata
some very similar results about decision problems for infinitary rational relations accepted by $2$-tape B\"uchi 
automata. 
\nl 
These very surprising results  provide the first examples of highly undecidable problems about the behaviour of very 
simple finite machines like $1$-counter automata or $2$-tape 
automata. 

\hs The paper is organized as follows. In Section 2 we recall some notions about arithmetical and analytical hierarchies and also about the Borel hierarchy. 
We study decision problems for infinite computations of $1$-counter automata in Section 3.  We infer some corresponding results about 
infinite computations of $2$-tape automata in Section 4. Some concluding remarks are given in Section 5. 

\section{Arithmetical and analytical hierarchies}

\subsection{Hierarchies of sets of integers}

\noi 
 The set of natural numbers is denoted by $\mathbb{N}$ and the set of all total functions from $\mathbb{N}$ into $\mathbb{N}$ will 
be denoted by $\mathcal{F}$. 

\hs 
We assume the reader to be familiar with the arithmetical   hierarchy on subsets of  $\mathbb{N}$.   We now recall   the notions
 of analytical hierarchy and of complete sets for classes of this hierarchy which may be found in \cite{rog}; see also for instance 
\cite{Odifreddi1,Odifreddi2} for more 
recent textbooks on computability theory. 

\begin{Deff}
Let  $k, l >0$ be some  integers. $\Phi$ is a partial computable functional of $k$ function variables and $l$ number variables if there exists $z\in \mathbb{N}$ 
such that for any $(f_1, \ldots , f_k, x_1, \ldots , x_l) \in \mathcal{F}^k \times \mathbb{N}^l$, we have 
$$\Phi (f_1, \ldots , f_k, x_1, \ldots , x_l) = \tau_z^{f_1, \ldots , f_k}(x_1, \ldots , x_l),$$
\noi where the right hand side is the output of the Turing machine with index $z$ and oracles $f_1, \ldots , f_k$ over the input $(x_1, \ldots , x_l)$. 
For $k>0$ and $l=0$, $\Phi$ is a partial computable functional  if, for some $z$, 
$$\Phi (f_1, \ldots , f_k) = \tau_z^{f_1, \ldots , f_k}(0).$$
\noi The value $z$ is called the Gödel number or index for $\Phi$. 
\end{Deff}

\begin{Deff}
Let $k, l >0$ be some  integers and $R  \subseteq \mathcal{F}^k \times \mathbb{N}^l$. The relation $R$ is said to be a computable relation of $k$ 
function variables and $l$ number variables if its characteristic function is computable. 
\end{Deff}

\noi We now define analytical subsets of $\mathbb{N}^l$.

\begin{Deff}
A subset $R$ of $\mathbb{N}^l$ is analytical if it is computable or if there exists a computable set $S  \subseteq \mathcal{F}^m \times \mathbb{N}^n$, with 
$m\geq 0$ and $n\geq l$, such that 
$$R = \{ (x_1, \ldots , x_l) \mid (Q_1s_1)(Q_2s_2) \ldots (Q_{m+n-l}s_{m+n-l}) S(f_1, \ldots , f_m, x_1, \ldots , x_n) \}, $$
\noi where $Q_i$ is either $\fa$ or $\exists$ for $1 \leq i \leq m+n-l$, and where $s_1, \ldots , s_{m+n-l}$ are $f_1, \ldots , f_m, x_{l+1}, \ldots , x_n$ in 
some order. 
\nl The expression $(Q_1s_1)(Q_2s_2) \ldots (Q_{m+n-l}s_{m+n-l}) S(f_1, \ldots , f_m, x_1, \ldots , x_n)$ is called a predicate form for $R$. A
quantifier applying over a function variable is of type $1$, otherwise it is of type $0$. 
In a predicate form the (possibly empty) sequence of quantifiers, indexed by their type, is called the prefix of the form. The reduced prefix is the sequence of 
quantifiers obtained by suppressing the quantifiers of type $0$ from the prefix. 
\end{Deff}

\noi We can now distinguish the levels of the analytical hierarchy by considering the number of alternations in the reduced prefix. 

\begin{Deff}
For $n>0$, a $\Si^1_n$-prefix is one whose reduced prefix begins with $\exists^1$ and has $n-1$ alternations of quantifiers. 
A $\Si^1_0$-prefix is one whose reduced prefix is empty. 
For $n>0$, a $\Pi^1_n$-prefix is one whose reduced prefix begins with $\fa^1$ and has $n-1$ alternations of quantifiers. 
A $\Pi^1_0$-prefix is one whose reduced prefix is empty. 
\nl A predicate form is a $\Si^1_n$ ($\Pi^1_n$)-form if it has a  $\Si^1_n$ ($\Pi^1_n$)-prefix. 
The class of sets in some $\mathbb{N}^l$ which can be expressed in $\Si^1_n$-form (respectively, $\Pi^1_n$-form) is denoted by 
$\Si^1_n$   (respectively, $\Pi^1_n$). 
\nl The class $\Si^1_0 = \Pi^1_0$ is the class of arithmetical sets. 
\end{Deff}

\noi We now recall some well known results about the analytical hierarchy. 

\begin{Pro}
Let $R \subseteq \mathbb{N}^l$ for some integer $l$. Then $R$ is an analytical set iff there is some integer $n\geq 0$ such that 
$R \in \Si^1_n$ or $R \in \Pi^1_n$. 
\end{Pro}

\begin{The} For each integer $n\geq 1$, 
\noi 
\begin{enumerate}
\ite[(a)] $\Si^1_n\cup \Pi^1_n \subsetneq  \Si^1_{n +1}\cap \Pi^1_{n +1}$.
\ite[(b)] A set $R \subseteq \mathbb{N}^l$ is in the class $\Si^1_n$ iff its 
complement is in the class $\Pi^1_n$. 
\ite[(c)] $\Si^1_n - \Pi^1_n \neq \emptyset$ and $\Pi^1_n - \Si^1_n \neq \emptyset$.
\end{enumerate}
\end{The}

\noi  Transformations of prefixes  are often used, following the rules given by the next theorem. 

\begin{The}
For any predicate form with the given prefix, an equivalent predicate form with the new one can be obtained, following the 
allowed prefix transformations given below :
\noi 
\begin{enumerate}
\ite[(a)]  $\ldots \exists^0 \exists^0 \ldots \ra \ldots  \exists^0 \ldots, $ ~~~~~~~~~~~~~~ \nl $ \ldots \fa^0  \fa^0 \ldots  \ra \ldots  \fa^0 \ldots ; $
\ite[(b)]  $\ldots \exists^1 \exists^1 \ldots \ra \ldots  \exists^1 \ldots, $~~~~~~~~~~~~~~\nl  $ \ldots \fa^1  \fa^1 \ldots  \ra \ldots  \fa^1 \ldots ;$
\ite[(c)]  $\ldots \exists^0 ~~~\ldots  \ra \ldots  \exists^1 \ldots, $~~~~~~~~~~~~ \nl $ \ldots  \fa^0 ~~~\ldots  \ra \ldots   \fa^1 \ldots ; $
\ite[(d)]  $\ldots \exists^0 \fa^1 \ldots \ra \ldots \fa^1 \exists^0 \ldots$, ~~~~~~  \nl $\ldots \fa^0 \exists^1 \ldots  \ra \ldots \exists^1 \fa^0 \ldots ; $
\end{enumerate}
\end{The}

\noi We can now define the notion of 1-reduction and of    $\Si^1_n$-complete (respectively,           $\Pi^1_n$-complete) sets. 
Notice that we give the definition for subsets of  $\mathbb{N}$ but this can be easily extended to subsets of $\mathbb{N}^l$ for some integer $l$. 

\begin{Deff}
Given two sets $A,B \subseteq \mathbb{N}$ we say A is 1-reducible to B and write $A \leq_1 B$
if there exists a total computable injective  function f from      $\mathbb{N}$     to   $\mathbb{N}$        with $A = f ^{-1}[B]$. 
\end{Deff}

\begin{Deff}
A set $A \subseteq \mathbb{N}$ is said to be $\Si^1_n$-complete   (respectively,   $\Pi^1_n$-complete)  iff $A$ is a  $\Si^1_n$-set 
 (respectively,   $\Pi^1_n$-set) and for each $\Si^1_n$-set  (respectively,   $\Pi^1_n$-set) $B \subseteq \mathbb{N}$ it holds that 
$B \leq_1 A$. 
\end{Deff}

\noi For each integer $n\geq 1$ there exist some $\Si^1_n$-complete subset of $\mathbb{N}$. Such sets are  precisely defined in \cite{rog} or \cite{cc}. 

\begin{Not}
$U_n$ denotes a  $\Si^1_n$-complete subset of $\mathbb{N}$. The set $U_n^-=\mathbb{N}-U_n  \subseteq \mathbb{N}$  is 
a $\Pi^1_n$-complete set.
\end{Not}

\subsection{Hierarchies of sets of infinite words}

We assume now  the reader to be familiar with the theory of formal ($\om$)-languages  
\cite{Thomas90,Staiger97}.
We shall follow usual notations of formal language theory. 
\nl  When $\Si$ is a finite alphabet, a {\it non-empty finite word} over $\Si$ is any 
sequence $x=a_1\ldots a_k$, where $a_i\in\Sigma$ 
for $i=1,\ldots ,k$ , and  $k$ is an integer $\geq 1$. The {\it length}
 of $x$ is $k$, denoted by $|x|$.
 The {\it empty word} has no letter and is denoted by $\lambda$; its length is $0$. 
 $\Sis$  is the {\it set of finite words} (including the empty word) over $\Sigma$.
 \nl  The {\it first infinite ordinal} is $\om$.
 An $\om$-{\it word} over $\Si$ is an $\om$ -sequence $a_1 \ldots a_n \ldots$, where for all 
integers $ i\geq 1$, ~
$a_i \in\Sigma$.  When $\sigma$ is an $\om$-word over $\Si$, we write
 $\sigma =\sigma(1)\sigma(2)\ldots \sigma(n) \ldots $,  where for all $i$,~ $\sigma(i)\in \Si$,
and $\sigma[n]=\sigma(1)\sigma(2)\ldots \sigma(n)$  for all $n\geq 1$ and $\sigma[0]=\lambda$.
\nl 
 The usual concatenation product of two finite words $u$ and $v$ is 
denoted $u.v$ (and sometimes just $uv$). This product is extended to the product of a 
finite word $u$ and an $\om$-word $v$: the infinite word $u.v$ is then the $\om$-word such that:
\nl $(u.v)(k)=u(k)$  if $k\leq |u|$ , and 
 $(u.v)(k)=v(k-|u|)$  if $k>|u|$.
\nl   
 The {\it set of } $\om$-{\it words} over  the alphabet $\Si$ is denoted by $\Si^\om$.
An  $\om$-{\it language} over an alphabet $\Sigma$ is a subset of  $\Si^\om$.  The complement (in $\Sio$) of an 
$\om$-language $V \subseteq \Sio$ is $\Sio - V$, denoted $V^-$.

\hs  We assume now the reader to be familiar with basic notions of topology which
may be found in \cite{Moschovakis80,LescowThomas,Kechris94,Staiger97,PerrinPin}.
There is a natural metric on the set $\Sio$ of  infinite words 
over a finite alphabet 
$\Si$ containing at least two letters which is called the {\it prefix metric} and defined as follows. For $u, v \in \Sio$ and 
$u\neq v$ let $\delta(u, v)=2^{-l_{\mathrm{pref}(u,v)}}$ where $l_{\mathrm{pref}(u,v)}$ 
 is the first integer $n$
such that the $(n+1)^{st}$ letter of $u$ is different from the $(n+1)^{st}$ letter of $v$. 
This metric induces on $\Sio$ the usual  Cantor topology for which {\it open subsets} of 
$\Sio$ are in the form $W.\Si^\om$, where $W\subseteq \Sis$.
A set $L\subseteq \Si^\om$ is a {\it closed set} iff its complement $\Si^\om - L$ 
is an open set.
Define now the {\it Borel Hierarchy} of subsets of $\Si^\om$:

\begin{Deff}
For a non-null countable ordinal $\alpha$, the classes ${\bf \Si}^0_\alpha$
 and ${\bf \Pi}^0_\alpha$ of the Borel Hierarchy on the topological space $\Si^\om$ 
are defined as follows:
\nl ${\bf \Si}^0_1$ is the class of open subsets of $\Si^\om$, 
\nl  ${\bf \Pi}^0_1$ is the class of closed subsets of $\Si^\om$, 
\nl and for any countable ordinal $\alpha \geq 2$: 
\nl ${\bf \Si}^0_\alpha$ is the class of countable unions of subsets of $\Si^\om$ in 
$\bigcup_{\gamma <\alpha}{\bf \Pi}^0_\gamma$.
 \nl ${\bf \Pi}^0_\alpha$ is the class of countable intersections of subsets of $\Si^\om$ in 
$\bigcup_{\gamma <\alpha}{\bf \Si}^0_\gamma$.
\end{Deff}

\noi For 
a countable ordinal $\alpha$,  a subset of $\Si^\om$ is a Borel set of {\it rank} $\alpha$ iff 
it is in ${\bf \Si}^0_{\alpha}\cup {\bf \Pi}^0_{\alpha}$ but not in 
$\bigcup_{\gamma <\alpha}({\bf \Si}^0_\gamma \cup {\bf \Pi}^0_\gamma)$.

\hs    
There are also some subsets of $\Si^\om$ which are not Borel. 
In particular 
the class of Borel subsets of $\Si^\om$ is strictly included into 
the class  ${\bf \Si}^1_1$ of {\it analytic sets} which are 
obtained by projection of Borel sets.

\noi  We now define completeness with regard to reduction by continuous functions. 
For a countable ordinal  $\alpha\geq 1$, a set $F\subseteq \Si^\om$ is said to be 
a ${\bf \Si}^0_\alpha$  
(respectively,  ${\bf \Pi}^0_\alpha$, ${\bf \Si}^1_1$)-{\it complete set} 
iff for any set $E\subseteq Y^\om$  (with $Y$ a finite alphabet): 
 $E\in {\bf \Si}^0_\alpha$ (respectively,  $E\in {\bf \Pi}^0_\alpha$,  $E\in {\bf \Si}^1_1$) 
iff there exists a continuous function $f: Y^\om \ra \Si^\om$ such that $E = f^{-1}(F)$. 
 ${\bf \Si}^0_n$
 (respectively ${\bf \Pi}^0_n$)-complete sets, with $n$ an integer $\geq 1$, 
 are thoroughly characterized in \cite{Staiger86a}.

\hs We recall now the definition of the  arithmetical hierarchy of  \ol s which form the effective analogue to the 
hierarchy of Borel sets of finite ranks. 
\nl Let $X$ be a finite alphabet. An \ol~ $L\subseteq X^\om$  belongs to the class 
$\Si_n$ if and only if there exists a recursive relation 
$R_L\subseteq (\mathbb{N})^{n-1}\times X^\star$  such that
$$L = \{\sigma \in X^\om \mid \exists a_1\ldots Q_na_n  \quad (a_1,\ldots , a_{n-1}, 
\sigma[a_n+1])\in R_L \}$$

\noi where $Q_i$ is one of the quantifiers $\fa$ or $\exists$ 
(not necessarily in an alternating order). An \ol~ $L\subseteq X^\om$  belongs to the class 
$\Pi_n$ if and only if its complement $X^\om - L$  belongs to the class 
$\Si_n$.  The inclusion relations that hold  between the classes $\Si_n$ and $\Pi_n$ are 
the same as for the corresponding classes of the Borel hierarchy. 
 The classes $\Si_n$ and $\Pi_n$ are  included in the respective classes 
${\bf \Si}_n^0$ and ${\bf \Pi}_n^0$ of the Borel hierarchy, and cardinality arguments suffice to show that these inclusions are strict. 

\hs  As in the case of the Borel hierarchy, projections of arithmetical sets 
 lead 
beyond the arithmetical hierarchy, to the analytical hierarchy of \ol s. The first class 
of this hierarchy is the (lightface) class $\Si^1_1$ of {\it effective analytic sets} 
 which are obtained by projection of arithmetical sets.
\nl  In fact an \ol~ $L\subseteq X^\om$  is in the class $\Si_1^1$ iff it is the projection 
of an \ol~ over the alphabet $X\times \{0, 1\}$ which is in the class $\Pi_2$.  The (lightface)  class $\Pi_1^1$ of  {\it effective co-analytic sets} 
 is simply the class of complements of effective analytic sets. We denote as usual $\Delta_1^1 = \Si^1_1 \cap \Pi_1^1$.

\hs  The Borel ranks of  (lightface) $\Delta_1^1$ sets   are the (recursive) 
ordinals  $\gamma < \om_1^{\mathrm{CK}}$, where $ \om_1^{\mathrm{CK}}$
 is the first non-recursive ordinal, usually called the Church-Kleene ordinal.  
Moreover, for every non null  ordinal $\alpha < \om_1^{\mathrm{CK}}$, there exist some  
${\bf \Si}^0_\alpha$-complete and some  ${\bf \Pi}^0_\alpha$-complete sets in the class $\Delta_1^1$.

\section{Infinite computations of $1$-counter   automata}

\hs Recall  the notion of acceptance of infinite words by Turing machines considered  by Castro and Cucker in \cite{cc}. 

\begin{Deff}
A non deterministic Turing machine $\mathcal{M}$ is a $5$-tuple $\mathcal{M}=(Q, \Si, \Ga, \delta, q_0)$, where $Q$ is a finite set of states, 
$\Si$ is a finite input alphabet, $\Ga$ is a finite tape alphabet satisfying $\Si  \subseteq \Ga$, $q_0$ is the initial state, 
and $\delta$ is a mapping from $Q \times \Ga$ to subsets of $Q \times \Ga \times \{L, R, S\}$. A configuration of $\mathcal{M}$ is a triple 
$(q, \sigma, i)$, where $q\in Q$, $\sigma \in \Ga^\om$ and $i\in \mathbb{N}$. An infinite sequence of configurations $r=(q_i, \alpha_i, j_i)_{i\geq 1}$
is called a run of $\mathcal{M}$ on $w\in \Sio$ iff: 
\begin{enumerate}
\ite[(a)] $(q_1, \alpha_1, j_1)=(q_0, w, 1)$, and 
\ite[(b)] for each $i\geq 1$, $(q_i, \alpha_i, j_i) \vdash (q_{i+1}, \alpha_{i+1}, j_{i+1})$, 
\end{enumerate}
\noi where $\vdash$ is the transition relation of $\mathcal{M}$ defined as usual. The run $r$ is said to be complete if 
$(\fa n \geq 1) (\exists k \geq 1) (j_k \geq n)$. The run $r$ is said to be oscillating if $(\exists k \geq 1) (\fa n \geq 1) (\exists m \geq n) ( j_m=k)$. 

\end{Deff}

\begin{Deff}
Let $\mathcal{M}=(Q, \Si, \Ga, \delta, q_0)$ be a non deterministic Turing machine   and $F \subseteq Q$. The $\om$-language accepted by $(\mathcal{M}, F)$ is 
the set of $\om$-words $ \sigma \in \Sio$ such that there exists a complete non oscillating run $ r=(q_i, \alpha_i, j_i)_{i\geq 1}$
 of $\mathcal{M}$  on  $\sigma$ such that, for all $ i, q_i \in F.$
\end{Deff}

\noi The above acceptance condition is denoted $1'$-acceptance in  \cite{CG78b}. Another usual acceptance condition is the now called B\"uchi 
acceptance condition which is also denoted $2$-acceptance in  \cite{CG78b}. 
We just now  recall its definition. 

\begin{Deff}
Let $\mathcal{M}=(Q, \Si, \Ga, \delta, q_0)$ be a non deterministic Turing machine   
and $F \subseteq Q$. The $\om$-language B\"uchi accepted by $(\mathcal{M}, F)$ is 
the set of $\om$-words $ \sigma \in \Sio$ such that there exists a complete non oscillating run $ r=(q_i, \alpha_i, j_i)_{i\geq 1}$
 of $\mathcal{M}$  on  $\sigma$ and  infinitely many integers $i$ such that $q_i \in F.$
\end{Deff}

\noi Recall that Cohen and Gold proved in \cite[Theorem 8.6]{CG78b} that one can effectively construct, from a given non deterministic Turing machine, 
another equivalent (i.e., accepting the same $\om$-language) 
 non deterministic Turing machine, equipped with the same kind of  acceptance condition, and in which every run is complete non oscillating. 

\hs Cohen and Gold proved also in \cite[Theorem 8.2]{CG78b} that an $\om$-language is accepted by a non deterministic Turing machine with 
$1'$-acceptance condition iff it is accepted by a non deterministic Turing machine with B\"uchi acceptance condition. It is known that $\om$-languages 
accepted by non deterministic Turing machines with $1'$ or  B\"uchi  acceptance condition form the (lightface) class $\Si_1^1$ of  effective analytic sets, 
\cite{Staiger97}.

\hs We now recall the definition of $k$-counter B\"uchi automata which will be useful in the sequel. 

\begin{Deff} Let $k$ be an integer $\geq 1$. 
A  $k$-counter machine ($k$-CM) is a 4-tuple 
$\mathcal{M}$=$(K,\Si, \Delta, q_0)$,  where $K$ 
is a finite set of states, $\Sigma$ is a finite input alphabet, 
 $q_0\in K$ is the initial state, 
and  $\Delta \subseteq K \times ( \Si \cup \{\lambda\} ) \times \{0, 1\}^k \times K \times \{0, 1, -1\}^k$ is the transition relation. 
The $k$-counter machine $\mathcal{M}$ is said to be {\it real time} iff: 
$\Delta \subseteq K \times
  \Si \times \{0, 1\}^k \times K \times \{0, 1, -1\}^k$, 
 i.e. iff there is no  $\lambda$-transitions. 
\nl  
If  the machine $\mathcal{M}$ is in state $q$ and 
$c_i \in \mathbf{N}$ is the content of the $i^{th}$ counter 
 $\mathcal{C}$$_i$ then 
the  configuration (or global state)
 of $\mathcal{M}$ is the  $(k+1)$-tuple $(q, c_1, \ldots , c_k)$.

\hs For $a\in \Si \cup \{\lambda\}$, 
$q, q' \in K$ and $(c_1, \ldots , c_k) \in \mathbf{N}^k$ such 
that $c_j=0$ for $j\in E \subseteq  \{1, \ldots , k\}$ and $c_j >0$ for 
$j\notin E$, if 
$(q, a, i_1, \ldots , i_k, q', j_1, \ldots , j_k) \in \Delta$ where $i_j=0$ for $j\in E$ 
and $i_j=1$ for $j\notin E$, then we write:
$$a: (q, c_1, \ldots , c_k)\mapsto_{\mathcal{M}} (q', c_1+j_1, \ldots , c_k+j_k)$$

\noi Thus we see that the transition relation must satisfy:
 \nl if $(q, a, i_1, \ldots , i_k, q', j_1, \ldots , j_k)  \in    \Delta$ and  $i_m=0$ for 
 some $m\in \{1, \ldots , k\}$, then $j_m=0$ or $j_m=1$ (but $j_m$ may not be equal to $-1$).

\hs  
Let $\sigma =a_1a_2 \ldots a_n \ldots $ be an $\om$-word over $\Si$. 
An $\om$-sequence of configurations $r=(q_i, c_1^{i}, \ldots c_k^{i})_{i \geq 1}$ is called 
a run of $\mathcal{M}$ on $\sigma$, starting in configuration 
$(p, c_1, \ldots, c_k)$, iff:
\begin{enumerate}
\ite[(1)]  $(q_1, c_1^{1}, \ldots c_k^{1})=(p, c_1, \ldots, c_k)$

\ite[(2)]   for each $i\geq 1$, there  exists $b_i \in \Si \cup \{\lambda\}$ such that
 $b_i: (q_i, c_1^{i}, \ldots c_k^{i})\mapsto_{\mathcal{M}}  
(q_{i+1},  c_1^{i+1}, \ldots c_k^{i+1})$  
and such that either ~  $a_1a_2\ldots a_n\ldots =b_1b_2\ldots b_n\ldots$ 
\nl or ~  $b_1b_2\ldots b_n\ldots$ is a finite prefix of ~ $a_1a_2\ldots a_n\ldots$
\end{enumerate}
\noi The run $r$ is said to be complete when $a_1a_2\ldots a_n\ldots =b_1b_2\ldots b_n\ldots$ 
\nl 
For every such run, $\mathrm{In}(r)$ is the set of all states entered infinitely
 often during run $r$.
\nl
A complete run $r$ of $M$ on $\sigma$, starting in configuration $(q_0, 0, \ldots, 0)$,
 will be simply called ``a run of $M$ on $\sigma$".
\end{Deff}

\begin{Deff} A B\"uchi $k$-counter automaton  is a 5-tuple 
$\mathcal{M}$=$(K,\Si, \Delta, q_0, F)$, 
where $ \mathcal{M}'$=$(K,\Si, \Delta, q_0)$
is a $k$-counter machine and $F \subseteq K$ 
is the set of accepting  states.
The \ol~ accepted by $\mathcal{M}$ is 

\hs $L(\mathcal{M})$= $\{  \sigma\in\Si^\om \mid \mbox{  there exists a  run r
 of } \mathcal{M} \mbox{ on } \sigma \mbox{  such that } \mathrm{In}(r)
 \cap F \neq \emptyset \}$

\end{Deff}

\noi  The class of \ol s accepted by  B\"uchi $k$-counter automata  will be 
denoted ${\bf BCL}(k)_\om$.
 The class of \ol s accepted by {\it  real time} B\"uchi $k$-counter automata  will be 
denoted {\bf r}-${\bf BCL}(k)_\om$.

\hs   Remark that  $1$-counter automata   introduced above are equivalent to pushdown automata 
whose stack alphabet is in the form $\{Z_0, A\}$ where $Z_0$ is the bottom symbol which always 
remains at the bottom of the stack and appears only there and $A$ is another stack symbol. 
\nl The class ${\bf BCL}(1)_\om$ is  a strict subclass of the class ${\bf CFL}_\om$ of context free \ol s
accepted by B\"uchi pushdown automata.

\hs Using a standard construction exposed for instance in \cite{HopcroftMotwaniUllman2001} we can construct, from a B\"uchi 
Turing machine, an equivalent $2$-counter automaton accepting the same $\om$-language with a B\"uchi acceptance condition. 

\hs Notice that these constructions are effective and that they can be achieved in an injective way. 
So we can now state the following lemma. 

\begin{Lem}\label{H1}
There is an injective computable function 
$H_1$ from $\mathbb{N}$ into $\mathbb{N}$ satisfying the following property. 
\nl  If $\mathcal{M}_z$ is the non deterministic Turing machine (equipped with a $1'$-acceptance condition) of index $z$, 
and if $\mathcal{A}_{H_1(z)}$ is the $2$-counter automaton 
(equipped with a $2$-acceptance condition) of index $H_1(z)$, 
then these two machines accept the same $\om$-language, i.e. $L(\mathcal{M}_z)=L(\mathcal{A}_{H_1(z)})$. 
\end{Lem}

\hs We are now going to recall some constructions which were used in \cite{Fin-mscs06} in the study of topological properties of context-free 
$\om$-languages. 

\hs Let $\Si$ be an alphabet having at least two letters,  $E$ be a new letter not in 
$\Si$,  $S$ be an integer $\geq 1$, and $\theta_S: \Sio \ra (\Sigma \cup \{E\})^\om$ be the 
function defined, for all  $x \in \Sio$, by: 
$$ \theta_S(x)=x(1).E^{S}.x(2).E^{S^2}.x(3).E^{S^3}.x(4) \ldots 
x(n).E^{S^n}.x(n+1).E^{S^{n+1}} \ldots $$

\noi It is proved in \cite{Fin-mscs06} that if    $L \subseteq \Sio$ is an 
$\om$-language in the class  ${\bf BCL}(2)_\om$ and   $k=cardinal(\Si)+2$, $S=(3k)^3$, then one can construct effectively, from 
a    B\"uchi $2$-counter automaton  $\mathcal{B}$  accepting $L$,                a real time 
B\"uchi $8$-counter automaton $\mathcal{A}$ such that $L(\mathcal{A})=\theta_S(L)$, so $\theta_S(L)$ 
is in the class  {\bf r}-${\bf BCL}(8)_\om$. This construction can be made  injective. 
On the other hand, it is easy to see that $\theta_S(\Sio)^-=(\Sigma \cup \{E\})^\om - \theta_S(\Sio)$ is accepted 
by a real time B\"uchi $1$-counter automaton.  The class 
{\bf r}-${\bf BCL}(8)_\om$ is closed by finite union in an effective way, so 
$\theta_S(L) \cup    \theta_S(\Sio)^-$ is accepted by a real time B\"uchi $8$-counter automaton which can be effectively constructed 
from         $\mathcal{B}$. 
 Thus we get the following result: 

\begin{Lem}\label{H2}
There is an injective computable function 
$H_2$ from $\mathbb{N}$ into $\mathbb{N}$ satisfying the following property. 
\nl  If $\mathcal{B}_z$ is the B\"uchi $2$-counter automaton  (reading words over $\Si$) of index $z$, 
and if $\mathcal{A}_{H_2(z)}$ is the real time B\"uchi $8$-counter automaton 
 of index $H_2(z)$, 
then  $L(\mathcal{A}_{H_2(z)}) = \theta_S( L(\mathcal{B}_z) ) \cup \theta_S(\Sio)^-$. 
\end{Lem}

\noi Another coding has been used in \cite{Fin-mscs06} which we now recall. 
Let    $K = 2 \times 3 \times 5 \times 7 \times 11 \times 13 \times 17 \times 19 = 9699690$       be       the product of the eight first prime numbers. 
Then an $\om$-word $x\in \Gao$ is coded by the $\om$-word 
$$h_K(x)=A.0^K.x(1).B.0^{K^2}.A.0^{K^2}.x(2).B.0^{K^3}.A.0^{K^3}.x(3).B \ldots  
B.0^{K^n}.A.0^{K^n}.x(n).B \ldots  $$

\noi over the alphabet $\Ga \cup \{A, B, 0\}$, where $A, B, 0$ are new letters not in $\Ga$. 
It is proved in \cite{Fin-mscs06} that, from a  real time B\"uchi $8$-counter automaton $\mathcal{A}$ accepting $L(\mathcal{A}) \subseteq \Gao$, 
one can effectively construct (in an injective manner) a  B\"uchi $1$-counter automaton accepting the $\om$-language 
$h_K( L(\mathcal{A}) )$$ \cup h_K(\Ga^{\om})^-$. 

\hs Consider  now  the mapping 
 $\phi_K: (\Ga \cup\{A, B, 0\})^\om \ra (\Ga \cup\{A, B, F,  0\})^\om $ which is simply defined by:  
for all $x\in (\Ga \cup\{A, B, 0\})^\om$, 
$$\phi_K(x) = F^{K-1}.x(1).F^{K-1}.x(2)  
\ldots F^{K-1}.x(n). F^{K-1}.x(n+1).F^{K-1} \ldots$$

\noi Then the $\om$-language 
$\phi_K ( h_K( L(\mathcal{A}) )$$ \cup h_K(\Ga^{\om})^- )$ is accepted by   a  real time B\"uchi $1$-counter automaton which can be effectively 
constructed from the real time B\"uchi $8$-counter automaton $\mathcal{A}$. On the other hand   
it is easy to see that the $\om$-language $ (\Ga \cup\{A, B, F,  0\})^\om  - \phi_K( (\Ga \cup\{A, B, 0\})^\om )$ is $\om$-regular 
and to construct a B\"uchi automaton 
accepting it. Then one can effectively construct from $\mathcal{A}$ a real time B\"uchi $1$-counter automaton accepting the $\om$-language 
$\phi_K ( h_K( L(\mathcal{A}) )$$ \cup h_K(\Ga^{\om})^- ) \cup \phi_K( (\Ga \cup\{A, B, 0\})^\om )^-$. This can be done in an injective manner. 
So we can state the following lemma. 

\begin{Lem}\label{H3}
There is an injective computable function 
$H_3$ from $\mathbb{N}$ into $\mathbb{N}$ satisfying the following property. 
\nl  If $A_z$ is the real time B\"uchi $8$-counter automaton  (reading words over $\Ga$) of index $z$, 
and if $A_{H_3(z)}$ is the real time B\"uchi $1$-counter automaton 
 of index $H_3(z)$ (reading words over $\Ga \cup \{A, B, F, 0\}$), 
then : 
$$L(A_{H_3(z)}) =\phi_K ( h_K( L(\mathcal{A}_z) ) \cup h_K(\Ga^{\om})^- ) \cup \phi_K( (\Ga \cup\{A, B, 0\})^\om )^-$$ 
\end{Lem}

\noi In the sequel we shall consider, as in \cite{cc},  that $\Si$ contains only two letters and we denote these letters by $a$ and $b$ so $\Si=\{a, b\}$. 
 Then $\Ga=\Si \cup \{ E \}$ and we set  $\Omega=\Ga \cup \{A, B, F, 0\}=\{a, b, E, A, B, F, 0\}$. 

\hs From now on, we shall  denote $\mathcal{M}_z$ the non deterministic Turing machine of index $z$, (reading words over $\Si$), 
equipped with a $1'$-acceptance condition, 
and $\mathcal{C}_z$ the  real time B\"uchi $1$-counter automaton of index $z$ (reading words over $\Omega$).   

\hs We set $H=H_3 \circ H_2 \circ H_1$, where $H_1$, $H_2$, and $H_3$ are the computable functions from $\mathbb{N}$ into $\mathbb{N}$
described  above, the functions $H_1$, $H_2$ and $H_3$ being given by Lemmas   \ref{H1},  \ref{H2}, and  \ref{H3}. 
Thus $H$ is an injective  computable function 
from $\mathbb{N}$ into $\mathbb{N}$ and if $z$ is the index of a non deterministic Turing machine reading words over $\Si$ and equipped 
with a $1'$-acceptance condition, then $H(z)$ is the index of a non deterministic real time B\"uchi $1$-counter automaton reading words over
the alphabet $\Omega=\{a, b, E, A, B, F, 0\}$.

\hs Notice also that a run $r$ of a real time B\"uchi $1$-counter automaton may be easily coded by an infinite word over the alphabet $\{0, 1\}$. 
We can then identified $r$ with its code $\bar{r}\in \{0, 1\}^\om$. 
Then it is easy to see that ``$r$ is a run of $\mathcal{C}_z$ over the $\om$-word 
$\sigma \in \Omega^\om$" and ``$r$ is an accepting run" can be  expressed by  arithmetical formulas. 

\hs We can now state that the universality  problem for 
$\om$-languages of     real time B\"uchi $1$-counter automata is highly undecidable. 
  
\begin{The}\label{U}
The universality  problem for 
$\om$-languages of     real time B\"uchi $1$-counter automata is $\Pi_2^1$-complete, i.e. 
the set  $\{ z \in \mathbb{N} \mid  L(\mathcal{C}_z) = \Omega^\om \}$ is  $\Pi_2^1$-complete. 

\end{The}

\proo We prove first that  this  set is in the class $\Pi_2^1$.  It suffices, as in the case of Turing machines, to write that 
$L(\mathcal{C}_z) = \Omega^\om$ if and only if  
``$\fa$ $\sigma \in \Omega^\om$ $\exists r \in \{0,1\}^\om $ ~~($r$ is an accepting run of $\mathcal{C}_z$ over $\sigma$)".
The two quantifiers of type 1 are followed by an arithmetical formula. Thus $\{ z \in \mathbb{N} \mid  L(\mathcal{C}_z) = \Omega^\om \}$ is in 
the class  $\Pi_2^1$. 
\nl In order to prove completeness we shall use the corresponding result for Turing machines proved in \cite{cc}: the set 
$\{ z \in \mathbb{N} \mid  L(\mathcal{M}_z) = \Si^\om \}$ is $\Pi_2^1$-complete. 
Consider now the injective  computable function  $H$ from $\mathbb{N}$ into $\mathbb{N}$ defined above.  We are going to prove  that,  
for each integer 
$z \in  \mathbb{N}$, it holds that  
$$L(\mathcal{M}_z) = \Si^\om \mbox{ if and only if }L(\mathcal{C}_{H(z) })= \Omega^\om.$$ 
 \noi By Lemma \ref{H2}, for  each integer $z \in \mathbb{N}$,   if $A_{H_2 \circ H_1(z)}$ is the real time B\"uchi $8$-counter automaton 
 of index $H_2 \circ H_1(z)$, 
then : $L(A_{H_2 \circ H_1(z)}) = \theta_S( L(\mathcal{M}_z) ) \cup \theta_S(\Sio)^-$.  
Thus $ L(\mathcal{M}_z)=\Sio$ iff  $L(A_{H_2 \circ H_1(z)}) = (\Si \cup \{E\})^\om$. 
\nl Next applying  Lemma \ref{H3} we see that 
$$L(\mathcal{C}_{H_3 \circ H_2 \circ H_1(z)}) = 
\phi_K ( h_K( L(A_{H_2 \circ H_1(z)} ) ) \cup h_K(\Ga^{\om})^- ) \cup \phi_K( (\Ga \cup\{A, B, 0\})^\om )^-$$
\noi  Thus $L(\mathcal{C}_{H_3 \circ H_2 \circ H_1(z)}) =  \Omega^\om $
\nl $\leftrightarrow$ $\phi_K ( h_K( L(A_{H_2 \circ H_1(z)} ) ) \cup h_K(\Ga^{\om})^- ) = \phi_K( (\Ga \cup\{A, B, 0\})^\om )$
\nl $\leftrightarrow$ $h_K( L(A_{H_2 \circ H_1(z)} ) ) \cup h_K(\Ga^{\om})^-   = (\Ga \cup\{A, B, 0\})^\om $
\nl $\leftrightarrow$ $L(A_{H_2 \circ H_1(z)} ) = \Ga^{\om}$
\nl $\leftrightarrow$ $ L(\mathcal{M}_z)=\Sio$. 
\hs This shows that $\{ z \in \mathbb{N} \mid  L(\mathcal{M}_z) = \Si^\om \} \leq_1 \{ z \in \mathbb{N} \mid  L(\mathcal{C}_z) = \Omega^\om \}$. 
Thus this latter set is  $\Pi_2^1$-complete. 
\ep

\begin{Rem}
An easy coding can be used to show that the above result still holds if we replace the alphabet $\Omega$ by a two letter alphabet (or even by an 
alphabet containing $n$ letters for an integer $n\geq 2$). This will be true for all the results presented in this paper. 
\end{Rem}

\begin{Rem}
If we consider context-free languages accepted by B\"uchi pushdown  automata, it is easy to see that the universality  problem is still in the class $\Pi_2^1$. 
Then we can infer from Theorem \ref{U} the following corollary. 
\end{Rem}

\begin{Cor}
The universality  problem for context-free
$\om$-languages accepted by B\"uchi pushdown  automata is $\Pi_2^1$-complete.
\end{Cor}

\noi Using a similar method as in the proof of Theorem  \ref{U}, we can prove the following result: 

\begin{The} 
The cofiniteness   problem for 
$\om$-languages of     real time B\"uchi $1$-counter automata is $\Pi_2^1$-complete, i.e. the set 
  $\{ z \in \mathbb{N} \mid  L(\mathcal{C}_z)  \mbox{ is cofinite } \}$  is   $ \Pi_2^1 \mbox{-complete. }$
\end{The}

\proo We first prove that the set $\{ z \in \mathbb{N} \mid  L(\mathcal{C}_z)  \mbox{ is cofinite } \}$  is in the class 
$ \Pi_2^1$. We can reason as in the corresponding proof for Turing machines in \cite{cc}. 
Consider a recursive bijection $b: (\mathbb{N}^\star)^2 \ra \mathbb{N}^\star$ and its inverse $b^{-1}$. 
Now we can consider an infinite word over a finite alphabet $\Omega$ as a countably infinite family of infinite words over the same 
alphabet by considering, for any $\om$-word $\sigma \in \Omega^\om$, the family of $\om$-words 
$(\sigma_i)$ suh that for each $i \geq 1$, the $\om$-word $\sigma_i \in \Omega^\om$ is defined by $\sigma_i(j)= \sigma(b(i, j))$ for each $j\geq 1$. 
\nl We can now express that $L(\mathcal{C}_z)  \mbox{ is cofinite }$ by a formula : 
\nl `` $\fa$ $\sigma \in \Omega^\om$ $\exists r \in \{0,1\}^\om $ $\exists i \geq 1$ 
 [ if (all $\om$-words $\sigma_i$, $i\geq 1$, are distinct), then ($r$ is an accepting run of $\mathcal{C}_z$ over $\sigma_i$) ]".
\nl This is a $\Pi_2^1$-formula  because ``all $\om$-words $\sigma_i$ are distinct" can be expressed by the arithmetical formula :
$(\fa j, k \geq 1)  (\exists i \geq 1) ~ \sigma_j(i) \neq \sigma_k(i)$. 

\hs To prove that the set $\{ z \in \mathbb{N} \mid  L(\mathcal{C}_z)  \mbox{ is cofinite } \}$  is   $ \Pi_2^1 \mbox{-complete}$, it suffices to remark that 
$ L(\mathcal{M}_z)$ is cofinite if and only if  $L(\mathcal{C}_{H_3 \circ H_2 \circ H_1(z)})= $ is cofinite. Thus 
$$ \{ z \in \mathbb{N} \mid  L(\mathcal{M}_z)  \mbox{ is cofinite } \} \leq_1  \{ z \in \mathbb{N} \mid  L(\mathcal{C}_z)  \mbox{ is cofinite } \}  $$
\noi So the completeness follows from the fact, proved in \cite{cc},  that the set $ \{ z \in \mathbb{N} \mid  L(\mathcal{M}_z)  \mbox{ is cofinite } \}$ 
is  $\Pi_2^1$-complete. 
\ep 

\hs As for the universality problem, we obtain the same complexity when considering context-free $\om$-languages. 

\begin{Cor}
The cofiniteness  problem for context-free
$\om$-languages accepted by B\"uchi pushdown  automata is $\Pi_2^1$-complete.
\end{Cor}

\noi We now determine the exact complexities of 
 the inclusion and the equivalence problems for $\om$-languages of     real time B\"uchi $1$-counter automata. 

\begin{The}
\noi  The inclusion and the equivalence problems for $\om$-languages of     real time B\"uchi $1$-counter automata are 
also  $\Pi_2^1$-complete, i.e. : 
\begin{enumerate}
\ite $\{ (y, z) \in \mathbb{N}^2  \mid  L(\mathcal{C}_y) \subseteq L(\mathcal{C}_z)  \}$ is  $\Pi_2^1$-complete. 
\ite $\{ (y, z) \in \mathbb{N}^2  \mid  L(\mathcal{C}_y) = L(\mathcal{C}_z)  \}$ is  $\Pi_2^1$-complete. 
\end{enumerate}
\end{The}

\proo  We first prove that the set $\{ (y, z) \in \mathbb{N}^2  \mid  L(\mathcal{C}_y)  \subseteq L(\mathcal{C}_z)  \}$ is a $\Pi_2^1$-set. 
It suffices to remark that  ``$L(\mathcal{C}_y) \subseteq L(\mathcal{C}_z)$" can be expressed by the  $\Pi_2^1$-formula :
`` $\fa$ $\sigma \in \Omega^\om$ $\fa r \in \{0,1\}^\om $ $\exists r' \in \{0,1\}^\om $ 
 [ if ($r$ is an accepting run of $\mathcal{C}_y$ over $\sigma$), then ($r'$ is an accepting run of $\mathcal{C}_z$ over $\sigma$) ]".
\nl Then the set $\{ (y, z) \in \mathbb{N}^2  \mid  L(\mathcal{C}_y) = L(\mathcal{C}_z)  \}$ which is the intersection of the two sets 
$\{ (y, z) \in \mathbb{N}^2  \mid  L(\mathcal{C}_y)  \subseteq L(\mathcal{C}_z)  \}$ and 
$\{ (y, z) \in \mathbb{N}^2  \mid  L(\mathcal{C}_z)  \subseteq L(\mathcal{C}_y)  \}$ is also a  $\Pi_2^1$-set.
\hs To prove completeness we denote  $n_0$  the index of a real time B\"uchi $1$-counter automaton accepting the 
$\om$-language $\Omega^\om$. Then we consider the function $F : \mathbb{N} \ra \mathbb{N}^2$ defined by 
$F(z)=(n_0, z)$. This function is injective and computable and for all $z \in  \mathbb{N}$ it holds that 
$ L(\mathcal{C}_z) = \Omega^\om $ iff 
$F(z)=(n_0, z) \in \{ (y, z) \in \mathbb{N}^2  \mid  L(\mathcal{C}_y)  \subseteq L(\mathcal{C}_z)  \}$. Thus Theorem \ref{U} implies that 
$\{ (y, z) \in \mathbb{N}^2  \mid  L(\mathcal{C}_y) \subseteq L(\mathcal{C}_z)  \}$ is  $\Pi_2^1$-complete.
\nl In a similar way, we prove that the set $\{ (y, z) \in \mathbb{N}^2  \mid  L(\mathcal{C}_y) = L(\mathcal{C}_z)  \}$ is  $\Pi_2^1$-complete. 
\ep 

\hs As for the previous results we easily get the following corollaries. 

\begin{Cor}
The inclusion and the equivalence problems for context-free
$\om$-languages accepted by B\"uchi pushdown  automata are $\Pi_2^1$-complete.
\end{Cor}

\noi A natural question about $1$-counter $\om$-languages or context-free $\om$-languages is the following one : ``can we decide whether a given 
$1$-counter $\om$-language (respectively, context-free $\om$-language) is regular, i.e. accepted by a B\"uchi automaton ?". 
We can state the following result. 

\begin{The}\label{reg}
\noi  The regularity  problem for $\om$-languages of     real time B\"uchi $1$-counter automata is
 $\Pi_2^1$-complete, i.e. : 
  the set $\{  z \in \mathbb{N}  \mid   L(\mathcal{C}_z) \mbox{ is regular } \}$ is  $\Pi_2^1$-complete. 

\end{The}

\proo We first prove that the set $\{  z \in \mathbb{N}  \mid   L(\mathcal{C}_z) \mbox{ is regular } \}$ is  in the class $\Pi_2^1$. 
We denote $R_C$ the set of indices of real time B\"uchi $1$-counter automata such that no transition of these automata change the counter 
value. So the counter value of these automata is always zero and they can be seen simply as B\"uchi automata. The set $R_C$ is obviously 
recursive and we can express $`` L(\mathcal{C}_z) \mbox{ is regular } "$ by the formula : 
$\exists y [ ~y \in R_C \mbox{ and  } L(\mathcal{C}_z) = L(\mathcal{C}_y) ~] $. The existential quantification is of type 0 and we have already seen that 
$L(\mathcal{C}_z) = L(\mathcal{C}_y)$ can be expressed by a  $\Pi_2^1$-formula. This proves that 
the set $\{  z \in \mathbb{N}  \mid   L(\mathcal{C}_z) \mbox{ is regular } \}$ is in the class  $\Pi_2^1$. 

\hs In order to prove the completeness, we shall use the following result of \cite{cc}. 
The set $P_{recursive}=\{  z \in \mathbb{N}  \mid  \exists y ~ L(\mathcal{M}_z)^-= L(\mathcal{M}_y) \}$ is 
$\Pi_2^1$-complete. 

\hs In fact Castro and Cucker defined a injective computable  function  $\varphi : \mathbb{N}\ra  \mathbb{N}$ such that : 
\nl (1) if $z \in U_2^-$ then $L(\mathcal{M}_{\varphi (z)})=\Sio$ (and so $\varphi (z) \in P_{recursive}$), and 
\nl (2) if $z \in U_2$ then  $\varphi (z) \notin P_{recursive}$. 

\hs Similarly we shall consider the function $H \circ \varphi$ which  is an injective and computable function from $\mathbb{N}$ into $\mathbb{N}$. And we are 
going to show that : 
\nl (1) if $z \in U_2^-$ then $L(\mathcal{C}_{H \circ \varphi (z)})=\Omega^\om$,   and 
\nl (2) if $z \in U_2$ then $L(\mathcal{C}_{H \circ \varphi (z)})$ is not a regular $\om$-language. 

\hs We consider now two cases. 
\nl {\bf First case.} If  $z \in U_2^-$ then $L(\mathcal{M}_{\varphi (z)})=\Sio$ so  $L(\mathcal{C}_{H \circ \varphi (z)})=\Omega^\om$. Thus in this case 
$L(\mathcal{C}_{H \circ \varphi (z)})$ is a regular $\om$-language. 
\nl {\bf Second  case.} If $z \in U_2$ then  $\varphi (z) \notin P_{recursive}$, i.e. $L(\mathcal{M}_{\varphi (z)})^-$ is not accepted by any Turing machine 
with $1'$ (or B\"uchi) acceptance condition. It is then easy to see that $L(\mathcal{C}_{H \circ \varphi (z)})^-$ is not accepted by any Turing machine with 
$1'$ (or B\"uchi) acceptance condition.  
Indeed if we denote again $A_{H_2 \circ H_1 \circ \varphi ( z)}$  the real time B\"uchi $8$-counter automaton 
 of index $H_2 \circ H_1\circ \varphi (z)$, 
then : $L(A_{H_2 \circ H_1\circ \varphi (z)}) = \theta_S( L(\mathcal{M}_{\varphi (z)}) ) \cup \theta_S(\Sio)^-$.  
Thus $L(A_{H_2 \circ H_1\circ \varphi (z)}) ^-= \theta_S(\Sio)-\theta_S( L(\mathcal{M}_{\varphi (z)}) )=\theta_S(L(\mathcal{M}_{\varphi (z)})^-)$ 
is not accepted by any Turing machine with 
$1'$ (or B\"uchi) acceptance condition.  Next  we see that 
$$L(\mathcal{C}_{ H \circ \varphi (z)}) = 
\phi_K ( h_K( L(A_{H_2 \circ H_1 \circ \varphi  (z)} ) ) \cup h_K(\Ga^{\om})^- ) \cup \phi_K( (\Ga \cup\{A, B, 0\})^\om )^-$$
\noi so its complement 
$$L(\mathcal{C}_{ H \circ \varphi (z)})^-=\phi_K ( h_K( L(A_{H_2 \circ H_1 \circ \varphi  (z)} )^- ) )$$
\noi is not accepted by any Turing machine with 
$1'$ (or B\"uchi) acceptance condition. In particular $L(\mathcal{C}_{H \circ \varphi (z)})$ is not a regular $\om$-language because otherwise its complement 
would be also regular hence accepted by a Turing machine. 

\hs Finally, using the reduction $H \circ \varphi $, we have proved that :
$$U_2^- \leq_1  \{  z \in \mathbb{N}  \mid   L(\mathcal{C}_z) \mbox{ is regular } \}$$
\noi and this proves that $\{  z \in \mathbb{N}  \mid   L(\mathcal{C}_z) \mbox{ is regular } \}$ is $\Pi_2^1$-complete. 
\ep

\hs We have also the following result about context-free $\om$-languages. 

\begin{Cor}
\noi  The regularity  problem for context-free $\om$-languages  accepted by B\"uchi pushdown  automata is 
 $\Pi_2^1$-complete. 
 
\end{Cor}

\noi We consider now the complementability problem and the determinizability problems. 
The complementability problem is $\Pi_2^1$-complete for $\om$-languages of Turing machines, i. e. the set 
$P_{recursive}=\{  z \in \mathbb{N}  \mid  \exists y  ~ L(\mathcal{M}_z)^-= L(\mathcal{M}_y) \}$ is 
$\Pi_2^1$-complete, \cite{cc}. We are going to show that it is also $\Pi_2^1$-complete for $\om$-languages  of     real time B\"uchi $1$-counter automata
or of B\"uchi pushdown  automata. 
We show also that the determinizability problems for $\om$-languages of     real time B\"uchi $1$-counter automata, 
or of B\"uchi pushdown  automata, are  $\Pi_2^1$-complete. We denote  $D_C$ the set of indices of 
{\it deterministic} real time B\"uchi $1$-counter automata. We can now state  the following result: 

\begin{The}
\noi  The complementability  problem and the    determinizability problem     for $\om$-languages of     real time B\"uchi $1$-counter automata are
 $\Pi_2^1$-complete, i.e. : 
\begin{enumerate}
\ite 
  $ \{  z \in \mathbb{N}  \mid  \exists y  ~~ L(\mathcal{C}_z)^- = L(\mathcal{C}_y) \}$ is  $\Pi_2^1$-complete. 
\ite  $\{  z \in \mathbb{N}  \mid  \exists y \in D_C ~~  L(\mathcal{C}_z)=L(\mathcal{C}_y) \}$ is $\Pi_2^1$-complete. 

\end{enumerate}
\end{The}

\proo  We first show that all these problems are in the class  $\Pi_2^1$. 
It is easy to see that $\{  z \in \mathbb{N}  \mid  \exists y   L(\mathcal{C}_z)^- = L(\mathcal{C}_y) \}$  is in the class $\Pi_2^1$ because 
$L(\mathcal{C}_z)^- = L(\mathcal{C}_y)$ can be expressed by a $\Pi_2^1$-formula and  the quantification $\exists y$ is of type $0$. 
\nl On the other hand, it is easy to see that the set  $D_C$ is recursive. The formula 
 $\exists y \in D_C ~~  L(\mathcal{C}_z)=L(\mathcal{C}_y)$ can be 
written : ``$\exists y [ y \in D_C  \mbox{ and  } L(\mathcal{C}_z)=L(\mathcal{C}_y) ] $" and it can be expressed by a $\Pi_2^1$-formula because the 
quantification $\exists y $ is of  type $0$ and $L(\mathcal{C}_z)=L(\mathcal{C}_y)$ can be expressed by a $\Pi_2^1$-formula. Thus  
the set  $\{  z \in \mathbb{N}  \mid  \exists y \in D_C ~~  L(\mathcal{C}_z)=L(\mathcal{C}_y) \}$ is in the class  $\Pi_2^1$. 

\hs Consider now the reduction $H \circ \varphi$ already considered in the proof of Theorem \ref{reg}. We have seen that there are two cases. 
\nl {\bf First case.} If  $z \in U_2^-$ then $L(\mathcal{M}_{\varphi (z)})=\Sio$ so  $L(\mathcal{C}_{H \circ \varphi (z)})=\Omega^\om$. In this case 
$L(\mathcal{C}_{H \circ \varphi (z)})$ is obviously accepted by a deterministic real time B\"uchi $1$-counter automaton. Moreover  its complement is empty 
therefore it is also accepted by a real time B\"uchi $1$-counter automaton. 
\nl {\bf Second  case.} If $z \in U_2$ then  $\varphi (z) \notin P_{recursive}$, and 
$L(\mathcal{C}_{H \circ \varphi (z)})^-$ is not accepted by any Turing machine with 
$1'$ (or B\"uchi) acceptance condition.  In particular, $L(\mathcal{C}_{H \circ \varphi (z)})^-$  is not accepted by any 
real time B\"uchi $1$-counter automaton. And $L(\mathcal{C}_{H \circ \varphi (z)})$ can not be accepted 
by any deterministic real time B\"uchi $1$-counter automaton because otherwise it would be in the arithmetical class 
$\Pi_2$ (see \cite{Staiger97}) and its complement would be accepted by a  Turing machine with 
$1'$ (or B\"uchi) acceptance condition. 

\hs This proves that : 
$$U_2^- \leq_1  \{  z \in \mathbb{N}  \mid  \exists y  ~~ L(\mathcal{C}_z)^- = L(\mathcal{C}_y) \}$$
\noi and 
$$U_2^- \leq_1  \{  z \in \mathbb{N}  \mid  \exists y \in D_C ~~  L(\mathcal{C}_z)=L(\mathcal{C}_y) \}$$
\noi and this ends the proof. \ep

\hs In a similar manner we prove the 
following result about context-free $\om$-languages. 

\begin{Cor}
\noi   The complementability  problem and the    determinizability problem   
for context-free $\om$-languages  accepted by B\"uchi pushdown  automata are 
 $\Pi_2^1$-complete. 
\end{Cor}

\noi We investigate now the unambiguity problem for $\om$-languages accepted by real time B\"uchi $1$-counter automata or by 
B\"uchi pushdown  automata. Recall that a real time B\"uchi $1$-counter automaton $\mathcal{A}$, accepting infinite words over an alphabet $\Omega$, 
 is said to be non ambiguous iff for every $\om$-word $x\in \Omega^\om$ there is at most one accepting run of $\mathcal{A}$ on $x$. 
An $\om$-language $L(\mathcal{A})$, accepted by a real time B\"uchi $1$-counter automaton $\mathcal{A}$, is said to be non ambiguous iff  there exists a 
 non ambiguous real time B\"uchi $1$-counter automaton $\mathcal{B}$ such that $L(\mathcal{B})=L(\mathcal{A})$; in the other case the $\om$-language 
$L(\mathcal{A})$ is said to be inherently ambiguous (notice that the notion of ambiguity refer here to acceptance by real time B\"uchi $1$-counter automata). 
The definition is similar for $\om$-languages accepted by B\"uchi pushdown  automata. A context-free $\om$-language $L$  is said to be non ambiguous iff  
 there exists a non ambiguous B\"uchi pushdown  automaton accepting $L$. 
It has been proved in \cite{Fin03b} that one cannot decide whether a given context-free $\om$-language $L$  is non ambiguous. 
We now state the following result. 

\begin{The}\label{unamb}
\noi  The unambiguity  problem      for $\om$-languages of     real time B\"uchi $1$-counter automata is 
 $\Pi_2^1$-complete, i.e. :  
 $$\mbox{ The set } \{  z \in \mathbb{N}  \mid  L(\mathcal{C}_z) \mbox{ is non ambiguous  }\}\mbox{  is } \Pi_2^1\mbox{-complete.} $$
\end{The}

\proo   We can first express 
``$\mathcal{C}_z$ is non ambiguous"  by : 
$$``\fa \sigma \in  \Omega^\om  \fa r, r' \in \{0, 1\}^\om  [(  r \mbox{ and } r' \mbox{ are accepting runs of  } \mathcal{C}_z  \mbox{ on }  \sigma) \ra  r=r']"$$
\noi which is a $\Pi_1^1$-formula. 
 Then `` $L(\mathcal{C}_z) $ is non ambiguous" can be expressed by the following formula:  
 ``$\exists y [ L(\mathcal{C}_z) = L(\mathcal{C}_y)  \mbox{ and }  \mathcal{C}_y   \mbox{ is non ambiguous}]$". This is a $\Pi_2^1$-formula because 
$L(\mathcal{C}_z) = L(\mathcal{C}_y) $ can be expressed by a $\Pi_2^1$-formula, and the quantification  $\exists y$ is of type $0$. 
Thus the set $\{  z \in \mathbb{N}  \mid  L(\mathcal{C}_z) \mbox{ is non ambiguous  }\}$  is  a $\Pi_2^1$-set.

\hs To prove completeness we shall use the following  result proved  in \cite{Fink-Sim}. 
Let $L(\mathcal{A})$ be a context-free $\om$-language accepted by a B\"uchi pushdown  automaton $\mathcal{A}$ such that $L(\mathcal{A})$ 
is an analytic but non Borel set. Then the set of $\om$-words, 
which have $2^{\aleph_0}$ accepting runs by $\mathcal{A}$, has cardinality $2^{\aleph_0}$. 
In particular $L(\mathcal{A})$ has the maximum degree of ambiguity; it is said to be inherently ambiguous of degree $2^{\aleph_0}$ in \cite{Fin03b}. 

\hs We define the following simple operations over $\om$-languages. For two $\om$-words $x, x' \in \Sio$ the $\om$-word $x \otimes x'$ is defined by : 
for every integer $n\geq 1$ ~$(x \otimes x')(2n -1)=x(n)$ and $(x \otimes x')(2n)=x'(n)$. 
For two $\om$-languages $L, L' \subseteq \Sio$, the $\om$-language $L \otimes L' $ is defined by $L \otimes L' = \{ x \otimes x' \mid x\in L \mbox{ and } 
x'\in L' \}$. 

\hs We shall in the sequel  use the following construction. 
We  know that there is a simple example of ${\bf \Si}^1_1$-complete set $L \subseteq \Sio$  accepted by a $1$-counter automaton, hence by a Turing machine 
with $1'$ acceptance condition, see \cite{Fin03a}. 
Then it is easy to define an injective computable function $\theta$ from $\mathbb{N}$ into $\mathbb{N}$ such that, for every integer $z \in \mathbb{N}$, 
it holds that $L(\mathcal{M}_{\theta(z)}) = (L  \otimes   \Sio)  \cup     (\Sio    \otimes   L(\mathcal{M}_{z}))$.   

\hs We are going to use now the reduction $H$ already considered above to show that the universality problem for $\om$-languages 
of     real time B\"uchi $1$-counter automata is 
 $\Pi_2^1$-complete. We have seen that  
 $$L(\mathcal{M}_z) = \Si^\om \mbox{ if and only if }L(\mathcal{C}_{H(z) })= \Omega^\om$$ 
\noi and we can easily see that 
 $$L(\mathcal{M}_{\theta(z)}) = \Si^\om \mbox{ if and only if } L(\mathcal{M}_z)= \Si^\om$$
\noi because $L \neq \Sio$. 

\hs The reduction $H \circ \theta$  is an injective computable function  from $\mathbb{N}$ into $\mathbb{N}$. 
\nl We consider now two cases. 
\nl {\bf First case.}  $L(\mathcal{M}_z)= \Si^\om$. Then $L(\mathcal{M}_{\theta(z)}) = \Si^\om$ and 
$L(\mathcal{C}_{H \circ \theta (z) })= \Omega^\om$. In particular $L(\mathcal{C}_{H \circ \theta (z) })$ is accepted by a non ambiguous 
real time B\"uchi $1$-counter automaton. 
\nl {\bf Second case.} $L(\mathcal{M}_z) \neq  \Si^\om$. Then there is an $\om$-word $x \in \Sio$ such that $x \notin L(\mathcal{M}_z)$. But 
$L(\mathcal{M}_{\theta(z)}) = (L  \otimes   \Sio)  \cup     (\Sio    \otimes   L(\mathcal{M}_{z}))$ thus 
$\{ \sigma \in \Sio \mid \sigma \otimes x \in L(\mathcal{M}_{\theta(z)}) \} = L$ is a ${\bf \Si}^1_1$-complete set. 
This implies that $L(\mathcal{M}_{\theta(z)}) $ 
is not a Borel set because otherwise its section $ \{ \sigma \in \Sio \mid \sigma \otimes x \in L(\mathcal{M}_{\theta(z)}) \} $
 would be also Borel, \cite{Kechris94}. 
\nl Recall that $H=H_3 \circ H_2 \circ H_1$, where $H_1$, $H_2$, and $H_3$ are the computable functions from $\mathbb{N}$ into $\mathbb{N}$
defined  above. 
 If $A_{H_2 \circ H_1 \circ \theta (z)}$ is the real time B\"uchi $8$-counter automaton 
 of index $H_2 \circ H_1 \circ \theta (z)$, 
then it is easy to see that $L(A_{H_2 \circ H_1 \circ \theta (z)}) = \theta_S( L(\mathcal{M}_{\theta(z)}) ) \cup \theta_S(\Sio)^-$ is not Borel.  
Next, considering the mappings $h_K$ and $\phi_K$, we can easily successively see that 
\nl $h_K( L(A_{H_2 \circ H_1 \circ \theta (z)} ) ) \cup h_K(\Ga^{\om})^-$ is not a Borel set, 
\nl $\phi_K ( h_K( L(A_{H_2 \circ H_1 \circ \theta (z)} ) ) \cup h_K(\Ga^{\om})^- ) $  is not a Borel set, 
\nl  $L(\mathcal{C}_{H_3 \circ H_2 \circ H_1 \circ \theta (z)}) = 
\phi_K ( h_K( L(A_{H_2 \circ H_1 \circ \theta (z)} ) ) \cup h_K(\Ga^{\om})^- ) \cup \phi_K( (\Ga \cup\{A, B, 0\})^\om )^-$ is not a Borel set, i.e. the 
$\om$-language $L(\mathcal{C}_{H \circ \theta (z)}) $ is not a Borel set. 
\nl Thus in that case the $\om$-language $L(\mathcal{C}_{H \circ \theta (z)}) $ is inherently ambiguous 
(and it is even inherently ambiguous of degree $2^{\aleph_0}$) , \cite{Fin03b}.

\hs Finally, using the reduction $H \circ \theta $, we have proved that : 
 $$\{ z \in \mathbb{N} \mid  L(\mathcal{M}_z) = \Si^\om \} \leq_1 \{ z \in \mathbb{N} \mid  L(\mathcal{C}_z)   \mbox{ is non ambiguous  } \}$$ 
\noi Thus this latter set is  $\Pi_2^1$-complete. 
\ep 

\hs In a similar manner we prove the 
following result about context-free $\om$-languages. 

\begin{Cor}
\noi   The unambiguity  problem 
for context-free $\om$-languages  accepted by B\"uchi pushdown  automata is 
 $\Pi_2^1$-complete. 
\end{Cor}

\noi 
A fundamental result due to Landweber is that one can determine in an effective manner the topological complexity of regular $\om$-languages: one can decide whether
a given regular $\om$-language is in a given Borel class (recall that  all regular $\om$-languages belong to  the class ${\bf \Delta}^0_3$), \cite{Landweber69}.
The question naturally arises of a similar problem for other classes of languages, like $\om$-languages of real time B\"uchi $1$-counter automata. It is 
proved in \cite{Fin-mscs06} that $\om$-languages of real time B\"uchi $1$-counter automata have the same topological complexity as $\om$-languages 
of Turing machines. 
From the above proof we can now infer that the topological complexity of $\om$-languages of real time B\"uchi $1$-counter automata is  
highly undecidable.

\begin{The}\label{borel-hard}
\noi Let $\alpha$ be a countable ordinal. Then  
\begin{enumerate}
\ite $ \{  z \in \mathbb{N}  \mid  L(\mathcal{C}_z) \mbox{ is in the Borel class } {\bf \Si}^0_\alpha \}$ is  $\Pi_2^1$-hard. 
\ite  $ \{  z \in \mathbb{N}  \mid  L(\mathcal{C}_z) \mbox{ is in the Borel class } {\bf \Pi}^0_\alpha \}$ is  $\Pi_2^1$-hard. 
\ite  $ \{  z \in \mathbb{N}  \mid  L(\mathcal{C}_z) \mbox{ is a  Borel set } \}$ is  $\Pi_2^1$-hard. 
\end{enumerate}
\end{The}

\proo  We can use the same reduction $H \circ \theta $ as in the proof of Theorem \ref{unamb}. We have seen that there are two cases. 
\nl {\bf First case.}  $L(\mathcal{M}_z)= \Si^\om$. Then $L(\mathcal{M}_{\theta(z)}) = \Si^\om$ and 
$L(\mathcal{C}_{H \circ \theta (z) })= \Omega^\om$. In particular $L(\mathcal{C}_{H \circ \theta (z) })$ is an open and closed subset of $\Omega^\om$
and it belongs to all Borel classes $ {\bf \Si}^0_\alpha$ and  ${\bf \Pi}^0_\alpha$. 
\nl {\bf Second case.} $L(\mathcal{M}_z) \neq  \Si^\om$. Then we have seen that the $\om$-language 
$L(\mathcal{C}_{H \circ \theta (z)}) $ is not a Borel set. 

\hs Finally, using the reduction $H \circ \theta $, we have proved that : 
 $$\{ z \in \mathbb{N} \mid  L(\mathcal{M}_z) = \Si^\om \} \leq_1 
\{ z \in \mathbb{N} \mid  L(\mathcal{C}_z)  \mbox{ is in the Borel class } {\bf \Si}^0_\alpha  \}$$ 
 $$\{ z \in \mathbb{N} \mid  L(\mathcal{M}_z) = \Si^\om \} \leq_1 
\{ z \in \mathbb{N} \mid  L(\mathcal{C}_z)  \mbox{ is in the Borel class } {\bf \Pi}^0_\alpha  \}$$ 
 $$\{ z \in \mathbb{N} \mid  L(\mathcal{M}_z) = \Si^\om \} \leq_1 
\{ z \in \mathbb{N} \mid  L(\mathcal{C}_z)  \mbox{ is a Borel set }   \}$$ 

\noi And this ends the proof since $\{ z \in \mathbb{N} \mid  L(\mathcal{M}_z) = \Si^\om \}$ is  $\Pi_2^1$-complete. 
\ep 

\hs In the case of context-free $\om$-languages accepted by B\"uchi pushdown automata the corresponding problems have been shown to be 
undecidable, using the undecidability of the Post correspondence problem \cite{Fin01a,Fin03a}. We can prove as above that they are in fact highly undecidable. 

\begin{Cor}
\noi   Let $\alpha$ be a countable ordinal. The following    problems are $\Pi_2^1$-hard. 
\begin{enumerate}
\ite 
   ``Determine whether a given  
 context-free $\om$-language   is in the  Borel class ${\bf \Si}^0_\alpha$ 
(respectively,  ${\bf \Pi}^0_\alpha$)".  
\ite 
 ``Determine whether a given  
 context-free $\om$-language    is  a Borel set". 
\end{enumerate}
\end{Cor}

\begin{Rem}\label{borel}
If $\alpha$ is an ordinal smaller than the Church-Kleene ordinal $\om_1^{\mathrm{CK}}$, i.e. is a 
recursive ordinal, then there exists  a universal set for   ${\bf \Si}^0_\alpha$-subsets of 
 $X^\om$ which 
is in the class $\Delta_1^1$. This is a known fact of Effective Descriptive Set Theory which  is proved in detail in \cite{Fink-Lec2}. 
This means that there exists a $\Delta_1^1$-set $U_\alpha \subseteq 2^\om \times X^\om$ such that for every set $L \subseteq X^\om$, $L$ is in the class 
${\bf \Si}^0_\alpha$ iff there is an $\om$-word $x \in 2^\om$ such that $[ \fa y \in X^\om ~ y \in L \leftrightarrow (x, y) \in U_\alpha]$, i.e. such that $L$ is the 
section of $U_\alpha$ in $x$. The $\Delta_1^1$-set $U_\alpha \subseteq 2^\om \times X^\om$ is accepted by a Turing machine with $1'$ or 
 B\"uchi  acceptance condition. Then we can prove that $\{ z \in \mathbb{N} \mid  L(\mathcal{C}_z)  \mbox{ is in the Borel class } {\bf \Si}^0_\alpha  \}$
is in fact a $\Si_3^1$-set. 
Similarly the existence of a $\Delta_1^1$ universal set for  ${\bf \Pi}^0_\alpha$-subsets of 
 $X^\om$ implies that $\{ z \in \mathbb{N} \mid  L(\mathcal{C}_z)  \mbox{ is in the Borel class } {\bf \Pi}^0_\alpha  \}$
is in fact a $\Si_3^1$-set. Similar results hold for  context-free $\om$-languages accepted by B\"uchi pushdown automata. 
\end{Rem}

\noi We consider now the arithmetical complexity of $\om$-languages of     real time B\"uchi $1$-counter automata. 
Here we get the exact complexity of  highly undecidable problems. 

\begin{The}
\noi Let $n \geq 1$ be an integer. Then 
\begin{enumerate}
\ite $ \{  z \in \mathbb{N}  \mid  L(\mathcal{C}_z) \mbox{ is in the arithmetical class }  \Si_n \}$ is  $\Pi_2^1$-complete. 
\ite $ \{  z \in \mathbb{N}  \mid  L(\mathcal{C}_z) \mbox{ is in the arithmetical class }  \Pi_n \}$ is  $\Pi_2^1$-complete. 
\ite $ \{  z \in \mathbb{N}  \mid  L(\mathcal{C}_z) \mbox{ is a }  \Delta^1_1 \mbox{-set } \}$ is  $\Pi_2^1$-complete. 

\end{enumerate}
\end{The}

\proo Let $n \geq 1$ be an integer. 
We first prove that  
$$ \{  z \in \mathbb{N}  \mid  L(\mathcal{C}_z) \mbox{ is in the arithmetical class }  \Si_n \}$$ 
\noi is a $\Pi_2^1$-set. 
We are going to use the existence of a universal set $\mathcal{U}_{ \Si_n}\subseteq \mathbb{N} \times \Omega^\om$ for the class of 
$\Si_n$-subsets of $\Omega^\om$, \cite[p. 172]{Moschovakis80}. The set $\mathcal{U}_{ \Si_n}$ is a 
$\Si_n$-subset of $\mathbb{N} \times \Omega^\om$ (i.e.  $(n, x) \in \mathcal{U}_{ \Si_n}$ can be expressed by a $\Si^0_n$-formula)
and for any $L \subseteq \Omega^\om$, ~$L$ is a $\Si_n$-set iff there is an integer $n$ such that ~$[ \fa x \in \Omega^\om ~
x\in L  \leftrightarrow (n, x) \in \mathcal{U}_{ \Si_n}]$. 
\nl Then we can express ``$L(\mathcal{C}_z) \mbox{ is in the arithmetical class }  \Si_n$" by the formula ``$\exists n \in \mathbb{N}~ \fa x \in \Omega^\om~
[ x\in L(\mathcal{C}_z)  \leftrightarrow (n, x) \in \mathcal{U}_{ \Si_n}]$". The formula ``$[ x\in L(\mathcal{C}_z)  \leftrightarrow (n, x) \in \mathcal{U}_{ \Si_n}]$"
is a $\Delta_2^1$-formula and the first quantifier $\exists$ is of type $0$. Therefore 
``$L(\mathcal{C}_z) \mbox{ is in}$ $\mbox{ the arithmetical class }  \Si_n$" can be expressed by a $\Pi_2^1$-formula. 
\nl The case of the arithmetical class $\Pi_n$ is very similar since there exists also a universal set 
$\mathcal{U}_{ \Pi_n}\subseteq \mathbb{N} \times \Omega^\om$ for the class of 
$\Pi_n$-subsets of $\Omega^\om$, \cite{Moschovakis80}.
\nl We now prove that $ \{  z \in \mathbb{N}  \mid  L(\mathcal{C}_z) \mbox{ is a }  \Delta^1_1 \mbox{-set } \}$ is a  $\Pi_2^1$-set. We have already 
seen that the set $P_{recursive}=\{  z \in \mathbb{N}  \mid  \exists y ~ L(\mathcal{M}_z)^-= L(\mathcal{M}_y) \}$ is 
$\Pi_2^1$-complete, \cite{cc}. On the other hand, an  $\om$-language   $L\subseteq X^\om$ is in the class $\Si_1^1$
iff it is accepted by a non deterministic Turing machine
with a  $1'$ or  B\"uchi acceptance condition, \cite{Staiger97}. Thus 
$P_{recursive}= \{  z \in \mathbb{N}  \mid  L(\mathcal{M}_z) \mbox{ is a }  \Delta^1_1 \mbox{-set } \}$. 
In a similar manner, $ \{  z \in \mathbb{N}  \mid  L(\mathcal{C}_z) \mbox{ is a }  \Delta^1_1 \mbox{-set } \} = 
\{  z \in \mathbb{N}  \mid \exists y ~ L(\mathcal{C}_z)^-= L(\mathcal{M}_y) \}$,  and it is easily seen to be in the class $\Pi_2^1$.

\hs We now prove completeness for the three problems. 
We can again use the same reduction $H \circ \theta $ as in the proof of Theorem \ref{unamb}. We have seen that there are two cases. 
\nl {\bf First case.}  $L(\mathcal{M}_z)= \Si^\om$. Then $L(\mathcal{M}_{\theta(z)}) = \Si^\om$ and 
$L(\mathcal{C}_{H \circ \theta (z) })= \Omega^\om$. In particular, for every integer $n\geq 1$, the $\om$-language 
 $L(\mathcal{C}_{H \circ \theta (z) })$ is in the arithmetical classes 
$\Si_n$ and $\Pi_n$. 
\nl {\bf Second case.} $L(\mathcal{M}_z) \neq  \Si^\om$. Then we have seen that the $\om$-language 
$L(\mathcal{C}_{H \circ \theta (z)}) $ is not a Borel set. Thus it is not a (lightface) $\Delta_1^1$-set and it is not in any arithmetical class 
$\Si_n$ or $\Pi_n$. 

\hs Finally, using the reduction $H \circ \theta $, we have proved that : 
 $$\{ z \in \mathbb{N} \mid  L(\mathcal{M}_z) = \Si^\om \} \leq_1 
\{ z \in \mathbb{N} \mid  L(\mathcal{C}_z)  \mbox{ is in the arithmetical class }  \Si_n \}$$ 
 $$\{ z \in \mathbb{N} \mid  L(\mathcal{M}_z) = \Si^\om \} \leq_1 
\{ z \in \mathbb{N} \mid  L(\mathcal{C}_z) \mbox{ is in the arithmetical class }  \Pi_n \}$$ 
 $$\{ z \in \mathbb{N} \mid  L(\mathcal{M}_z) = \Si^\om \} \leq_1 
\{ z \in \mathbb{N} \mid  L(\mathcal{C}_z) \mbox{ is a }  \Delta^1_1 \mbox{-set }\}$$ 

\noi And this ends the proof since $\{ z \in \mathbb{N} \mid  L(\mathcal{M}_z) = \Si^\om \}$ is  $\Pi_2^1$-complete. 
\ep 

\hs In a similar way, we can prove the following result for context-free $\om$-languages accepted by B\"uchi pushdown automata. 
Notice that the decision problems cited in the following corollary were shown to be undecidable in \cite{Fin01a,Fin03a} but their exact (high) complexity 
was unexpected. 

\begin{Cor}
\noi   Let $n \geq 1$ be an integer. The following decision   problems are $\Pi_2^1$-complete. 
\begin{enumerate}
\ite 
   ``Determine whether a given  
 context-free $\om$-language   is in the   arithmetical class $\Si_n$ (respectively, $\Pi_n$)"
\ite 
 ``Determine whether a given  
 context-free $\om$-language    is  a $\Delta^1_1$-set". 
\end{enumerate}
\end{Cor}

\section{Infinite computations of $2$-tape  automata}

\noi We are going to study now  decision problems  about the infinite behaviour of 
$2$-tape B\"uchi automata  accepting  infinitary rational relations. 
We first recall the definition of $2$-tape B\"uchi automata and of  infinitary rational relations. 

\begin{Deff}
A  2-tape B\"uchi automaton 
 is a sextuple $\mathcal{T}=(K, \Si_1, \Si_2, \Delta, q_0, F)$, where 
$K$ is a finite set of states, $\Si_1$ and $\Si_2$ are finite  alphabets, 
$\Delta$ is a finite subset of $K \times \Si_1^\star \times \Si_2^\star \times K$ called 
the set of transitions, $q_0$ is the initial state,  and $F \subseteq K$ is the set of 
accepting states. 
\nl A computation $\mathcal{C}$ of the  
2-tape B\"uchi automaton $\mathcal{T}$ is an infinite sequence of transitions 
$$(q_0, u_1, v_1, q_1), (q_1, u_2, v_2, q_2), \ldots (q_{i-1}, u_{i}, v_{i}, q_{i}), 
(q_i, u_{i+1}, v_{i+1}, q_{i+1}), \ldots $$
\noi The computation is said to be successful iff there exists a final state $q_f \in F$ 
and infinitely many integers $i\geq 0$ such that $q_i=q_f$. 
\nl The input word of the computation is $u=u_1.u_2.u_3 \ldots$
\nl The output word of the computation is $v=v_1.v_2.v_3 \ldots$
\nl Then the input and the output words may be finite or infinite. 
\nl The infinitary rational relation $R(\mathcal{T})\subseteq \Si_1^\om \times \Si_2^\om$ 
accepted by the 2-tape B\"uchi automaton $\mathcal{T}$ 
is the set of couples $(u, v) \in \Si_1^\om \times \Si_2^\om$ such that $u$ and $v$ are the input 
and the output words of some successful computation $\mathcal{C}$ of $\mathcal{T}$. 
\nl The set of infinitary rational relations will be denoted ${\bf RAT}_\om$. 
\end{Deff} 

\noi In order to prove that some decision problems about the infinite behaviour of 
$2$-tape B\"uchi automata are highly undecidable, we shall use the results of the preceding section and a coding used in a 
previous paper on the topological complexity of infinitary rational relations. 
We proved in \cite{Fin06b} that infinitary rational relations have the same topological complexity as $\om$-languages 
accepted by B\"uchi Turing machines. This very surprising result was obtained by using a simulation of the behaviour of real time 
$1$-counter automata by $2$-tape B\"uchi automata. 
We recall now a coding which was used in \cite{Fin06b}. 

\hs  We now first  define a coding of  an $\om$-word over the finite alphabet $\Omega=\{a, b, E, A, B, F, 0\}$
by an  $\om$-word over the  alphabet $\Omega' = \Omega \cup \{C\}$, where  $C$ is an additionnal letter 
not in $\Omega$. 

\hs For $x\in \Omega^\om$  the $\om$-word $h(x)$ is defined by : 
$$h(x) = C.0.x(1).C.0^2.x(2).C.0^3.x(3).C \ldots C.0^n.x(n).C.0^{n+1}.x(n+1).C \ldots$$
\noi Then it is easy to see that the mapping $h$ from $\Omega^\om$ into $(\Omega \cup \{C\})^\om$ is continuous and injective. 

\hs Let now  $\alpha$ be the $\om$-word over the alphabet $\Omega'$ 
 which is simply defined by:
$$\alpha = C.0.C.0^2.C.0^3.C.0^4.C \ldots C.0^n.C.0^{n+1}.C \ldots$$

\noi The following results were proved in \cite{Fin06b}.

\begin{Lem}\label{R1}
Let $\Omega$  be a finite alphabet such that $0\in \Omega$, 
$\alpha$ be the $\om$-word over  $\Omega \cup \{C\}$ defined as above, and 
 $L \subseteq \Omega^\om$ be in  {\bf r}-${\bf BCL}(1)_\om$.
Then there exists  an infinitary rational relation 
$R_1 \subseteq  (\Omega \cup \{C\})^\om \times (\Omega \cup \{C\})^\om$ such that:
$$\fa x\in \Omega^{\om}~~~ (x\in L) \mbox{  iff } ( (h(x), \alpha) \in R_1 )$$ 
\end{Lem}

\begin{Lem}\label{complement} The set 
$R_2 = (\Omega \cup \{C\})^\om \times (\Omega \cup \{C\})^\om - ( h(\Omega^{\om}) \times \{\alpha\} )$
 is an infinitary rational relation.  
\end{Lem}

\noi Considering the union $R_1 \cup R_2$ of the two infinitary rational relations obtained in the two above lemmas we get the following result. 

\begin{Pro}  Let  $L \subseteq \Omega^\om$ be in  {\bf r}-${\bf BCL}(1)_\om$ and $\mathcal{L}= h(L)  \cup (h(\Omega^{\om}))^- $.   Then 
 $$R = \mathcal{L} \times \{\alpha\} ~~ \bigcup  ~~(\Omega')^\om \times ( (\Omega')^\om - \{\alpha\})$$
\noi is an  infinitary rational relation. 
\end{Pro}

\noi Moreover it is proved in \cite{Fin06b} that one can construct effectively, from a real time $1$-counter B\"uchi automaton $\mathcal{A}$ accepting $L$, 
a $2$-tape B\"uchi automaton $\mathcal{B}$ accepting the infinitary relation 
$R = \mathcal{L} \times \{\alpha\} ~~ \bigcup  ~~(\Omega')^\om \times ( (\Omega')^\om - \{\alpha\})$. 
\nl This can be done in an injective way, so we get the following result. 

\hs Notice that from now on we shall denote $\mathcal{T}_z$ the 
$2$-tape B\"uchi automaton of index $z$. 

\begin{Lem}  
There is an injective computable function 
$H'$ from $\mathbb{N}$ into $\mathbb{N}$ satisfying the following property. 
\nl  If $\mathcal{C}_z$ is the real time B\"uchi $1$-counter automaton (reading words over $\Omega$)
 of index $z$, 
and if $\mathcal{T}_{H'(z)}$ is the $2$-tape B\"uchi automaton 
 of index $H'(z)$, 
then : $R(\mathcal{T}_{H'(z)}) = 
( h( L(\mathcal{C}_z) )  \cup (h(\Omega^{\om}))^- )\times \{\alpha\} ~~ \bigcup  ~~(\Omega')^\om \times ( (\Omega')^\om - \{\alpha\})$. 
\end{Lem}

\noi We can now state our first results about $2$-tape B\"uchi automata. Notice that the four   decision problems considered here  
were known to be undecidable. But the proof used the undecidability of Post correspondence problem, 
as in the case of finitary rational relations stated in \cite{Berstel79}, in such a way that 
these decision problems were 
only proved to be hard for the first level of the {\it arithmetical} hierarchy. We obtain here the exact complexity of these problems which is surprisingly 
high.

\begin{The}\label{U2}
The universality  problem, the cofiniteness problem, the equivalence problem,  and the inclusion problem  for  infinitary rational relations 
 are  $\Pi_2^1$-complete, i.e. : 
\begin{enumerate}
\ite  $\{ z \in \mathbb{N} \mid  R(\mathcal{T}_z) = \Omega'^\om  \times \Omega'^\om \}$ is  $\Pi_2^1$-complete. 
\ite  $\{ z \in \mathbb{N} \mid  R(\mathcal{T}_z)  \mbox{ is cofinite } \}$ is  $\Pi_2^1$-complete.
\ite $\{ (y, z) \in \mathbb{N}^2  \mid  R(\mathcal{T}_y) \subseteq R(\mathcal{T}_z)  \}$ is  $\Pi_2^1$-complete. 
\ite $\{ (y, z) \in \mathbb{N}^2  \mid  R(\mathcal{T}_y) = R(\mathcal{T}_z)  \}$ is  $\Pi_2^1$-complete. 

\end{enumerate}
\end{The}

\proo  In order to prove that these problems are in the class $\Pi_2^1$, we can reason as in the case of  $\om$-languages of 
real time B\"uchi $1$-counter automata. 

\hs To prove completeness, we use the reduction $H'$ defined above and the following properties which can be  easily checked. For each integer $z$, 
\begin{enumerate}
\ite $L(\mathcal{C}_z)=\Omega^\om $ iff  $R(\mathcal{T}_{H'(z)}) = \Omega'^\om \times \Omega'^\om$. 
\ite $L(\mathcal{C}_z)$ is cofinite iff $R(\mathcal{T}_{H'(z)}) $ is cofinite.
\ite $L(\mathcal{C}_y) \subseteq L(\mathcal{C}_z) $ iff  $R(\mathcal{T}_{H'(y)}) \subseteq R(\mathcal{T}_{H'(z)}) $.
\ite $L(\mathcal{C}_y) = L(\mathcal{C}_z) $ iff  $R(\mathcal{T}_{H'(y)}) = R(\mathcal{T}_{H'(z)}) $.
\end{enumerate}

\noi Then the completeness results follow easily from the corresponding results about  $\om$-languages of 
real time B\"uchi $1$-counter automata, proved in the preceding section. 

\ep 

\hs We consider now the  ``regularity problem" for infinitary rational relation. 
An infinitary rational relation $R\subseteq \Si_1^\om \times \Si_2^\om$ may be seen
as an $\om$-language over the product alphabet $\Si_1 \times \Si_2$. Then a relation $R \subseteq  \Si_1^\om \times \Si_2^\om$
 is accepted by a B\"uchi automaton iff 
it is accepted by a $2$-tape B\"uchi automaton with two reading heads which move synchronously. The relation $R$ is then called a synchronized infinitary 
rational relation. These relations have been studied by  Frougny and Sakarovitch   in   \cite{FrougnySakarovitch93} where they proved that one cannot decide 
whether a given   infinitary rational relation is synchronized. We shall prove that actually this problem is also $\Pi_2^1$-complete. This is also the case for the 
complementability  problem, the    determinizability problem, and the unambiguity  problem for infinitary rational relations. 
We denote below $T_D$ the (recursive) set of  indices of {\it deterministic} $2$-tape B\"uchi automata. 

\begin{The}
The  ``regularity problem",  the 
complementability  problem, the    determinizability problem, and the unambiguity  problem for  infinitary rational relations 
 are  $\Pi_2^1$-complete, i.e. : 
\begin{enumerate}
\ite  $\{ z \in \mathbb{N} \mid  R(\mathcal{T}_z)  \mbox{ is a synchronized  rational relation }  \}$ is  $\Pi_2^1$-complete. 
\ite $\{ z \in \mathbb{N} \mid R(\mathcal{T}_z)^-  \mbox{ is an infinitary   rational relation }  \}$ is  $\Pi_2^1$-complete.  
\ite  $\{ z \in \mathbb{N} \mid  \exists y \in T_D ~ R(\mathcal{T}_z) =  R(\mathcal{T}_y)   \}$ is  $\Pi_2^1$-complete.  
\ite $\{ z \in \mathbb{N} \mid  R(\mathcal{T}_z)  \mbox{ is a non ambiguous  rational relation }  \}$ is  $\Pi_2^1$-complete. 
\end{enumerate}
\end{The}

\proo We can reason as in the case of  $\om$-languages of 
real time B\"uchi $1$-counter automata to prove that these problems are in the class $\Pi_2^1$.  

\hs To prove completeness we consider the reduction $H \circ \theta$ already used in the proof of Theorem \ref{unamb}. And we shall use now the reduction 
$H' \circ H \circ \theta$, where $H'$ is defined above in this section. 
The  reduction $H' \circ H \circ \theta$  is an injective computable function  from $\mathbb{N}$ into $\mathbb{N}$. Returning to the proof of Theorem 
 \ref{unamb}, we can see that there are now two cases. 

\hs  {\bf First case.}  $L(\mathcal{M}_z)= \Si^\om$. Then $L(\mathcal{M}_{\theta(z)}) = \Si^\om$ and 
$L(\mathcal{C}_{H \circ \theta (z) })= \Omega^\om$ and $R(\mathcal{T}_{H' \circ H \circ \theta(z)})=\Omega'^\om \times \Omega'^\om$. 
Thus in that case $R(\mathcal{T}_{H' \circ H \circ \theta(z)})$ is a synchronized  rational relation accepted by a 
deterministic, hence also non ambiguous,  $2$-tape B\"uchi automaton. And its complement is empty so it is also an  
 infinitary rational relation. 
\nl {\bf Second case.} $L(\mathcal{M}_z) \neq  \Si^\om$. Then we have seen that in that case  the 
$\om$-language $L(\mathcal{C}_{H \circ \theta (z)}) $ is not a Borel set. It is easy to see that the 
 infinitary rational relation $R(\mathcal{T}_{H' \circ H \circ \theta(z)})$ is also a non Borel set. 
\nl Thus in that case $R(\mathcal{T}_{H' \circ H \circ \theta(z)})$ is not a synchronized  rational relation because otherwise it would be a 
${\bf \Delta}^0_3$-set. The relation $R(\mathcal{T}_{H' \circ H \circ \theta(z)})$ 
 can not be accepted by any  
deterministic  $2$-tape B\"uchi automaton because otherwise it would be a ${\bf \Pi}^0_2$-set.  
The relation   $R(\mathcal{T}_{H' \circ H \circ \theta(z)})$
 is  inherently ambiguous 
(and it is even inherently ambiguous of degree $2^{\aleph_0}$, see \cite{Fin03b, Fink-Sim}). And the complement 
$\Omega'^\om \times \Omega'^\om - R(\mathcal{T}_{H' \circ H \circ \theta(z)})$ is not an analytic set 
(because otherwise $R(\mathcal{T}_{H' \circ H \circ \theta(z)})$ would be analytic and coanalytic hence Borel). Thus the complement of 
$R(\mathcal{T}_{H' \circ H \circ \theta(z)})$ is not an  infinitary rational relation.

\hs Finally, using the reduction $H' \circ H \circ \theta $, we have proved that : 
 $\{ z \in \mathbb{N} \mid  L(\mathcal{M}_z) = \Si^\om \}$ is reduced to the four problems we consider here. Thus these problems are 
$\Pi_2^1$-complete. 
\ep 

\hs Topological and arithmetical properties of infinitary rational relations have been shown to be undecidable in \cite{Finkel03e}. The proofs used the 
undecidability of Post correspondence problem and the existence of an analytic but non Borel set proved in \cite{Finkel03d}. So classical decision problems were 
only proved to be hard for the first level of the {\it arithmetical} hierarchy. 

\hs 
 We can now infer from the proof of the preceding theorem, reasoning as in the case of  $\om$-languages of 
real time B\"uchi $1$-counter automata,   that topological and arithmetical properties of infinitary rational relations are actually highly undecidable.

\begin{The}
\noi Let $\alpha$ be a non null countable ordinal. Then  
\begin{enumerate}
\ite $ \{  z \in \mathbb{N}  \mid  R(\mathcal{T}_z) \mbox{ is in the Borel class } {\bf \Si}^0_\alpha \}$ is  $\Pi_2^1$-hard. 
\ite  $ \{  z \in \mathbb{N}  \mid  R(\mathcal{T}_z) \mbox{ is in the Borel class } {\bf \Pi}^0_\alpha \}$ is  $\Pi_2^1$-hard. 
\ite  $ \{  z \in \mathbb{N}  \mid  R(\mathcal{T}_z) \mbox{ is a  Borel set } \}$ is  $\Pi_2^1$-hard. 
\end{enumerate}
\end{The}

\begin{The}
\noi Let $n \geq 1$ be an integer. Then 
\begin{enumerate}
\ite $ \{  z \in \mathbb{N}  \mid  R(\mathcal{T}_z) \mbox{ is in the arithmetical class }  \Si_n \}$ is  $\Pi_2^1$-complete. 
\ite $ \{  z \in \mathbb{N}  \mid  R(\mathcal{T}_z) \mbox{ is in the arithmetical class }  \Pi_n \}$ is  $\Pi_2^1$-complete. 
\ite $ \{  z \in \mathbb{N}  \mid  R(\mathcal{T}_z) \mbox{ is a }  \Delta^1_1 \mbox{-set } \}$ is  $\Pi_2^1$-complete. 

\end{enumerate}
\end{The}

\section{Concluding remarks and further work} 

\noi We have got very surprising results which show  that many decision problems about $\om$-languages of 
real time B\"uchi $1$-counter automata   and infinitary rational relations exhibit actually a great complexity, despite the simplicity 
of the definition of $1$-counter automata or $2$-tape automata. 
\nl Recall that, by Remark \ref{borel},  if  $\alpha$ is an ordinal smaller than the Church-Kleene ordinal $\om_1^{\mathrm{CK}}$, 
then  $\{ z \in \mathbb{N} \mid  L(\mathcal{C}_z)  \mbox{ is in the Borel class } {\bf \Si}^0_\alpha  \}$
and    $\{ z \in \mathbb{N} \mid  L(\mathcal{C}_z)  \mbox{ is in}$ $\mbox{the Borel class } {\bf \Pi}^0_\alpha  \}$ are   $\Si_3^1$-sets. 
Moreover they are $\Pi_2^1$-hard by Theorem \ref{borel-hard}. However the exact complexity of being in the Borel class ${\bf \Si}^0_\alpha $ (respectively, 
${\bf \Pi}^0_\alpha $), for a countable ordinal $\alpha$,  remains an open problem for $\om$-languages of real time $1$-counter automata 
(respectively, pushdown automata,  $2$-tape automata). 

\hs May be one of the most surprising results in this paper  is that the universality problem for infinitary rational relations accepted by 
$2$-tape B\"uchi automata is $\Pi_2^1$-complete. This result may be compared to the complexity of the universality problem for timed 
B\"uchi automata. Alur and Dill proved in \cite{AlurDill94} that the universality problem for timed 
B\"uchi automata is $\Pi_1^1$-hard. On the other hand this problem is known to be in the class $\Pi_2^1$  but its exact 
complexity is still unknown. 
Notice that using the $\Pi_1^1$-hardness of the universality problem for timed 
B\"uchi automata some other decision problems for timed 
B\"uchi automata have been shown to be $\Pi_1^1$-hard, \cite{AlurDill94,Fin06}.

\hs  Recognizable languages of  infinite bidimensional words (infinite pictures)  have been recently studied in \cite{ATW02,Finkel04}. 
Using partly  similar reasoning as in this paper we have  proved 
 that some decision problems for recognizable languages of infinite pictures have the same degrees 
as the corresponding problems about $\om$-languages of real time $1$-counter automata, \cite{Fink-tilings}.  
Notice that some problems, like  the non-emptiness problem and the infiniteness problem,   are 
$\Si^1_1$-complete for  recognizable languages of infinite pictures   but are decidable for  $\om$-languages of real time $1$-counter automata or 
$2$-tape automata. Some problems studied in  \cite{Fink-tilings} are specific to  languages of infinite pictures. In particular,  
 it is  $\Pi_2^1$-complete to determine whether a given B\"uchi 
recognizable language of infinite pictures can be accepted row by row using an automaton model over ordinal words of length 
$\om^2$.

\end{document}